\documentclass[12pt]{article}

\usepackage{times}

\usepackage{graphicx}
\usepackage{epstopdf}
\usepackage{subfigure}
\usepackage{graphicx}
\usepackage[small]{caption}
\usepackage{amsmath}
\usepackage{footnote}
\usepackage{color}
\usepackage{multirow}
\usepackage{amsmath}

\title{Modeling and Simulation of the Economics of Mining in the Bitcoin Market}

\author{Luisanna Cocco \and Michele Marchesi 
}

\author
{Luisanna Cocco,$^{1\ast}$ Michele Marchesi,$^{1}$\\
\\
\normalsize{$^{1}$Department of electrical and electronic engineering,
University of Cagliari,}\\
\normalsize{Piazza D'Armi, 09123 Cagliari, Italy.}
\\
\normalsize{$^\ast$ E-mail: \{luisanna.cocco, michele\}@diee.unica.it}
} 

\date{}

\begin{document} 


\baselineskip16pt


\maketitle 

\begin{abstract}
In January 3, 2009, Satoshi Nakamoto gave rise to the "Bitcoin Block Chain" creating the first block of the chain hashing on his computer’s central processing unit (CPU). Since then, the hash calculations to mine Bitcoin have been getting more and more complex, and consequently the mining hardware evolved to adapt to this increasing difficulty.
Three generations of mining hardware have followed the CPU's generation. They are GPU's, FPGA's and ASIC's generations.

This work presents an agent based artificial market model of the Bitcoin mining process and of the Bitcoin transactions. The goal of this work is to model the economy of the mining process, starting from GPU's generation, the first with economic significance. 

The model reproduces some "stylized facts" found in real time price series and some core aspects of the mining business.
In particular, the computational experiments performed are able to reproduce the unit root property, the fat tail phenomenon and the volatility clustering of Bitcoin price series.
In addition, under proper assumptions, they are able to reproduce the price peak at the end of November 2013, its next fall in April 2014, the generation of Bitcoins, the hashing capability, the power consumption, and the mining hardware and electrical energy expenses of the Bitcoin network.

\textbf{keywords}: Artificial Financial Market, Bitcoin, Heterogeneous Agents, Market Simulation
\end{abstract}

\section{Introduction}

Bitcoin is a digital currency alternative to the legal ones, as any other crypto currency.
Nowadays, Bitcoin is the most popular cryptocurrency.
It was created by a cryptologist known as "Satoshi Nakamoto", whose real identity is still unknown \cite{Satoshi}.
Like other cryptocurrencies, Bitcoin uses cryptographic techniques and, thanks to an open source system, anyone is allowed to inspect and even modify the source code of the Bitcoin software.

The Bitcoin network is a peer-to-peer network that monitors and manages both the generation of new Bitcoins and the consistency verification of transactions in Bitcoins. This network is composed of a high number of computers connected to each other through the Internet. They perform complex cryptographic procedures which generate new Bitcoins (mining) and manage the Bitcoin transactions register, verifying their correctness and truthfulness.

Mining is the process which allows to find the so called "proof of work" that validates a set of transactions and adds them to the massive and transparent ledger of every past Bitcoin transaction known as the "Blockchain".
The generation of Bitcoins is the reward for the validation process of the transactions.
The Blockchain was generated starting since January 3, 2009 by the inventor of the Bitcoin system himself, Satoshi Nakamoto. The first block is called "genesis block" and contains a single transaction, which generates 50-Bitcoin for the benefit of the creator of the block.
The whole system is set up to yield just 21 million Bitcoins by 2040, and over time the process of mining will become less and less profitable. The main source of remuneration for the miners in the future will be the fees on transactions, and not the mining process itself.

In this work, we propose an agent-based model with the aim to study and analyse the mining process and the Bitcoin market starting from September 1, 2010, the approximate date when miners started to buy mining hardware to mine Bitcoins. 
The proposed model simulates the mining process and the Bitcoin transactions, by implementing a mechanism for the formation of the Bitcoin price, and specific behaviors for each typology of trader. 
We try to reproduce the generation of Bitcoins, the main stylized facts present in the real Bitcoin market and the economy of the mining process.
The model described is built on a previous work of the authors \cite{Cocco2014}, which modeled the Bitcoin market under a purely financial perspective. In this work, we fully consider also the economics of mining.

The paper is organized as follows. In Section \textit{Related Work} we discuss other works related to this paper, in Section \textit{Mining Process} we describe briefly the mining process and we give an overview on the mining hardware and on its evolution over time. In Section \textit{The Model} we present the proposed model in detail. Section \textit{Simulation Results} presents the values given to several parameters of the model and reports the results of the simulations, including an analysis of Bitcoin real prices, and a robustness analysis. The conclusions of the paper are reported in Section \textit{Conclusions}. Finally, appendix deals with the calibration to some parameters of the model. 

\section{Related Work}\label{sec:2}

The study and analysis of the cryptocurrency market is a relatively new field. 
In the last years, several papers appeared on this topic given its potential interest and the many issues related to it (see for instance the works \cite{Androulaki,Bergstra,Brezo,Eyal,Hanley,Hout,Moore,Singh}).

However, very few works were made to model the cryptocurrencies market. We can cite the works by Luther \cite{Luther}, who studied why some cryptocurrencies failed to gain widespread acceptance using a simple agent model; by  Bornholdt et al. \cite{Bornholdt}, who proposed a model based on Moran process to study the cryptocurrencies able to emerge; by Garcia et al. \cite{Garcia}, who studied the role of social interactions in the creation of price bubbles; by Kristoufek \cite{Kristoufek} who analysed the main drivers of the Bitcoin price; and by Kaminsky et al. \cite{Kaminsky} who related the Bitcoin market with its sentiment analysis on social networks.

In this paper we propose a complex agent-based model in order to reproduce the economy of the mining process and the main stylized facts of the Bitcoin price series. 
Our model is inspired by business, economic and financial agent-based models that depict how organizations, or in general the economy of a country, create, deliver, and capture value.

As regards the business models, Amini et al. \cite{Amini} presented a agent-based model with the aim to analyze the impact of alternative production and sales policies on the diffusion of a new product; Cocco et al. \cite{Cocco2013,Cocco2011,Cocco2014bis} proposed agent-based models to simulate the software market and analyze the business processes and policies adopted by proprietary software firms and Open Source software firms; Li et al. \cite{Li} researched the dominant players’ behavior in supply chains and the relationship between the selling prices and purchasing prices in supply chains by using a multi-agent simulation model; Rohitratana et al. \cite{Rohitratana} studied the pricing schemes of the market of the Software as-a-Service and on the market of the proprietary or traditional software; finally, Xiaoming et al. \cite{Xiaoming} studied how a firm maximizes its profit by determining the production and sales policies for a new product during the lifetime of the product. 

Concerning economic models, in \cite{EURACE} the authors presented one of the most significant agent-based model developed to date in order to study the European economy. In particular, they show how monetary policies, i.e, credit money supplied by commercial banks as loans to firms, influence the economy of a country.
In \cite{Dosi2010,Dosi2013} agent based keynesian models are presented in order to investigate the properties of macro economic dynamics and the impact of public polices on supply, demand and the fundamentals of the economy, and to study the interactions between income distribution and monetary and fiscal policies.

As regards artificial financial market models, they reproduce the real functioning of markets, trying to explain the main stylised facts observed in financial markets, such as the fat-tailed distribution of returns, the volatility clustering, and the unit-root property. For a review, see works \cite{Chakra} and \cite{Chen}.
Raberto et al. \cite{Raberto2001} and Cincotti et al. \cite{Cincotti} proposed the Genoa Artificial Stock Market (GASM) an agent-based artificial financial market characterized by actual tracking of status and wealth of each agent, and by a realistic trading and price clearing mechanisms. GASM is able to reproduce some of the main stylised facts observed in real financial markets.

This paper is built on GASM, adding specific features and a mix of zero-intelligence and trend-following traders with the aim to model the Bitcoin exchange market and the economics of mining.

\section{Mining Process}\label{sec:3}

Today, every few minutes thousands of people send and receive Bitcoins through  the peer-to-peer electronic cash system created by Satoshi Nakamoto. All transactions are public and stored in a distributed database called Blockchain which is used to confirm transactions and prevent the double-spending problem.

People who confirm transactions of Bitcoins and store them in the Blockchain are called "miners".
As soon as new transactions are notified to the network, miners check their validity and authenticity and collect them in a block. Then, they take the information contained in the block of the transactions, which include a variable number called "nonce" and run the SHA-256 hashing algorithm on this block, turning the initial information into a sequence of 256 bits, known as Hash \cite{CourtoisGrajek}.

There is no way of knowing how this sequence will look before calculating it, and the introduction of a minor change in the initial data causes a drastic change in the resulting Hash.  

The miners cannot change the data containing the information of transactions, but can change the "nonce" number used to create a different hash. The goal is to find a Hash having a given number of leading zero bits. This number can be varied to change the difficulty of the problem.
The first miner who creates a proper Hash with success (he finds the "proof-of-work"), gets a reward in Bitcoins, and the successful Hash is stored with the block of the validated transactions in the Blockchain. 

In a nutshell, 

\begin{quote}
\small{"Bitcoin miners make money when they find a 32-bit value which,
when hashed together with the data from other transactions with a standard
hash function gives a hash with a certain number of 60 or more zeros. This is an extremely rare event}", \cite{CourtoisGrajek}.
\end{quote}

The steps to run the network are the followings:

\begin{quote}
\small{" New transactions are broadcast to all nodes; each node collects new transactions into a block; each node works on finding a difficult proof-of-work for its block; when a node finds a proof-of-work, it broadcasts the block to all nodes; nodes accept the block only if all transactions in it are valid and not already spent; nodes express their acceptance of the block by working on creating the next block in the chain, using the hash of the accepted block as the previous hash"}, \cite{Satoshi}.
\end{quote}

Producing a single hash is computationally very easy, consequently in order to regulate the generation of Bitcoins, over time the Bitcoin protocol makes this task more and more difficult.
 
The proof-of-work is implemented by incrementing the nonce in the block until a value is found that gives the block's hash with the required leading zero bits. If the hash does not match the required format, a new nonce is generated  and the Hash calculation starts again \cite{Satoshi}.
Countless attempts may be necessary before finding a nonce able to generate a  correct Hash.

The computational complexity of the process necessary to find the proof-of-work is adjusted over time in such a way that the number of blocks found each day is more or less constant (approximately 2016 blocks in two weeks, one every 10 minutes).
In the beginning, each generated block corresponded to the creation of 50 Bitcoins, this number being halved each four years, after 210,000 blocks additions. So, the miners have a reward equal to 50 Bitcoins if the created blocks belong to the first 210,000 blocks of the Blockchain, 25 Bitcoins if the created blocks range from the 210,001th to the 420,000th block in the Blockchain, 12.5 Bitcoins if the created blocks range from the 420,001th to the 630,000th block in the Blockchain, and so on.

Over time, mining Bitcoin is getting more and more complex, due to the increasing number of miners, and the increasing power of their hardware.
We have witnessed the succession of four generations of hardware, i.e. CPU's, GPU's, FPGA's and ASIC's generation, each of them characterized by a specific hash rate (measured in H/sec) and power consumption. 
With time, the power and the price of the mining hardware has been steadly increasing, though the price of H/sec has been decreasing. To face the increasing costs, miners are pooling together to share resources.

\subsection{The evolution of the mining hardware}\label{sec:3.1}
In January 3, 2009, Satoshi Nakamoto created the first block of the Blockchain, called "Genesis Block", hashing on the central processing unit (CPU) of his computer. 
Like him, the early miners mined Bitcoin running the software on their personal computers. 
The CPU's era represents the first phase of the mining process, the other eras being GPU's, FPGA's and ASIC's eras (see web site https://tradeblock.com/blog/ \\the-evolution-of-mining/).

Each era announces the use of a specific typology of mining hardware. 
In the second era, started about on September 2010, boards based on graphics processing unit (GPU) running in parallel entered the market, giving rise to the GPU era. 

About in December 2011, the FPGA's era started and hardware based on field programmable gate array cards (FPGA) specifically designed to mine Bitcoins was available in the market. Finally, in 2013 fully customized application specific integrated circuit (ASIC) appeared, substantially increasing the hashing capability of the Bitcoin network and marking the beginning of the fourth era.

Over time, the different mining hardware available was characterized by an increasing hash rate, a decreasing power consumption per hash, and increasing costs.
For example, NVIDIA Quadro NVS 3100M, 16 cores, belonging to the GPU generation, has a hash rate equal to 3.6 MH/s and a power consumption equal to 14 W \cite{Courtois}; ModMiner Quad, belonging to the FPGA generation, has a hash rate equal to 800 MH/s and power consumption equal to 40 W \cite{Courtois}; Monarch(300), belonging to the ASIC generation, has a hash rate equal to 300 GH/s and power consumption equal to 175 W (see web site https://tradeblock.com/mining/. 

\subsection{Modeling the Mining Hardware Performances}\label{sec:3.2}
The goal of our work is to model the economy of the mining process, so we  neglected the first era, when Bitcoins had no monetary value, and miners used the power available on their PCs, at almost no cost. We simulated only the remaining three generations of mining hardware. 

We gathered  information about the products that entered the market in each era to model these three generations of hardware, in particular
with the aim to compute:

\begin{itemize}
\item the average hash rate per US\$ spent on hardware, $R(t)$, expressed in $\frac{H}{sec * \$}$;
\item the average power consumption per $H/sec$, $P(t)$, expressed in $\frac{W}{H/sec}$.
\end{itemize}

The average hash rate and the average power consumption were computed averaging the real market data at specific times and constructing two fitting curves. 

To calculate the hash rate and the power consumption of the mining hardware of the GPU era, that we estimate ranging from September 1st, 2010 to September 29th, 2011, we computed an average for $R$ and $P$ taking into account some representative products in the market during that period, neglecting the costs of the motherboard.

In that era, motherboards with more than one Peripheral Component Interconnect express (PCIe), started to enter the market allowing to install, by using adapters, multiple video cards in only one system and to mine criptocurrency, thanks to the power of the GPUs.
In Table \ref{tab:GPU}, we describe the features of some GPUs in the market in that period. The data reported are taken from the web site http://coinpolice.com/gpu/. 

\begin{table}%
\caption{\textit{GPU Mining Hardware}.\label{tab:GPU}}{%
\begin{tabular}{|l|l|l|l|}
\hline    
Date&Product&Hash Rate GH/\$&Consumption W/GH\\
\hline
\multirow{4}{*}{23/09/2009 } & Radeon 5830 & 0.001475	& 593.22\\
&Radeon 5850 & 0.0015 & 398.94\\
&Radeon 5870 & 0.0015 & 467.66\\
&Radeon 5970 & 0.0023 & 392\\
\hline
\multirow{3}{*}{22/10/2010 } & Radeon 6870 & 0.0015 & 503.33\\
&Radeon 6950 & 0.002 &	500\\
&Radeon 6990 & 0.0018 &	328.95\\
\hline
\end{tabular}}
\end{table}

As regards the FPGA and ASIC eras, starting about on September 2011 and on December 2013, respectively, we tracked the history of the mining hardware by following the introduction into the market of Butterfly Labs company's products.
We extracted the data illustrated in Table \ref{prodotti} from the history of the web site \textit{http://www.butterflylabs.com/} through the web site \textit{web.archive.org.}. For hardware in the market in 2014 and 2015 we referred to the Bitmain Technologies Ltd company, and in particular, to the mining hardware called AntMiner (see web site https://bitmaintech.com and Table \ref{prodotti}).

\begin{table}
\caption{\textit{Butterfly Labs Mining Hardware}: FPGA Hardware from 09/29/2011 to 12/17/2012, ASIC Hardware from 12/17/2012 to December 2013 and AntMiner Hardware for 2014 and 2015.\label{prodotti}}{%
\scalebox{0.63}{
\begin{tabular}{|l|l|l|l|l|l|}
\hline  
Date&Product &Price \$&Hash Rate GH/s&Hash Rate $\frac{GH}{sec*\$}$&Power Consumption $\frac{W}{GH/sec}$\\
\hline
09/29/2011- 12/2/2011&The Single&699&1&0.0014&19.8\\
\hline
\multirow{2}{*}{12/2/2011- 12/28/2011}&The Single&699&1&0.0014&19.8\\
&Rig Box&24980&50.4&0.0021&49\\
\hline
\multirow{2}{*}{12/28/2011- 05/1/2012}&The Single&599&0.832&0.0014&96.15\\
&Rig Box&24980&50.4&0.0021&49\\
\hline
\multirow{2}{*}{05/1/2012- 12/17/2012}&The Single&599&0.832&0.0014&96.15\\
&Mini Rig &15295&25.2&0.0016&49\\
\hline
\multirow{2}{*}{12/17/2012- 04/10/2013 }&BitForce Jalapeno&149&4.5&0.0302&1\\
&BitForce Little Single SC&649&30&0.0462&1\\
&BitForce Single SC&1299&60&0.0462&1\\
&BitForce Mini Rig SC&29899&1500&0.0502&1\\
\hline
\multirow{2}{*}{04/10/2013- 05/31/2013 }&Bitcoin Miner&274&5&0.0182&6\\
&Bitcoin Miner&1249&25&0.02&6\\
&Bitcoin Miner&2499&50&0.02&6\\
\hline
\multirow{2}{*}{ 05/31/2013- 10/15/2013 }&Bitcoin Miner&274&5&0.0182&6\\
&Bitcoin Miner&1249&25&0.02&6\\
&Bitcoin Miner&2499&50&0.02&6\\
&Bitcoin Miner&22484&500&0.0222&6\\
\hline
\multirow{2}{*}{ 10/15/2013- 12/10/2013}&Bitcoin Miner&274&5&0.0182&6\\
&Bitcoin Miner&2499&50&0.02&6\\
&Bitcoin Miner&22484&500&0.0222&6\\
&Bitcoin Minin Card&2800&300&0.1071&0.6\\
&Bitcoin Minin Card&4680&600&0.1282&0.6\\
\hline
12/10/2013- 01/22/2014&AntminerS1&734.18&180&0.245&2\\
\hline
01/22/2014- 07/4/2014&AntminerS2&1715&1000&0.583&1.1\\
\hline
07/4/2014- 10/23/2014&AntminerS4-B2&1250&2000&1.6&0.69\\
\hline
10/23/2014- 03/25/2015&AntminerS5-B5&419&1155&2.756&0.51\\
\hline
03/25/2015-30/09/2015&AntminerS7-B8&454&4730&10.42&0.27\\
\hline
\end{tabular}}}
\end{table}

Starting from the mining products in each period (see Tables \ref{tab:GPU} and \ref{prodotti}), we fitted a "best hash rate per \$" and a "best power consumption function" (see Table \ref{average}). We call the fitting curves $R(t)$ and $P(t)$, respectively.

\begin{table}%
\caption{\textit{Average of Hash Rate  and of Power Consumption over time. }\label{average}}{%
\scalebox{0.82}{
\begin{tabular}{|l|l|l|}
\hline  
Date $\Rightarrow$ Simulation Step&Average of Hash Rate $\frac{GH}{sec*\$}$&Average of power Consumption $\frac{W}{GH/sec}$\\
\hline
September 1, 2010 $\Rightarrow$ 1&0.0017&454.87\\
\hline
September 29, 2011 $\Rightarrow$ 394&0.0014&19.8\\
\hline
December 2,2011 $\Rightarrow$ 458&$0.00175$&34.4\\
\hline
December 28,2011 $\Rightarrow$ 484&$0.0017$&72.575\\
\hline
May 1, 2012 $\Rightarrow$ 608&$0.0029$&72.575\\
\hline
December 17, 2012 $\Rightarrow$ 835&0.03565&1\\
\hline
 April 10, 2013 $\Rightarrow$ 953&0.0194&6\\
\hline
May 31, 2013 $\Rightarrow$ 1004&0.0201&6\\
\hline
October 15, 2013 $\Rightarrow$ 1141&0.1351&3.84\\
\hline
December 10, 2013 $\Rightarrow$ 1197&0.0595&3.84\\
\hline
January 22, 2014 $\Rightarrow$ 1240&0.245&2\\
\hline
July 4, 2014 $\Rightarrow$ 1403&0.583&1.1\\
\hline
October 23, 2014 $\Rightarrow$ 1484&1.6&0.69\\
\hline
March 25, 2015 $\Rightarrow$ 1667&2.756&0.51\\
\hline
September 30, 2015 $\Rightarrow$ 1856&10.42&0.27\\
\hline
\end{tabular}}}
\end{table}

We used a general exponential model to fit the curve of the hash rate, $R(t)$ obtained by using eq. \ref{R}:

\begin{equation}\label{R}
R(t) =    a*e^{(b*t)}
\end{equation}

where  $a= 8.635*10^4$ and $b=0.006318$.

The fitting curve of the power consumption $P(t)$ is also a general exponential model:

\begin{equation}\label{P}
P(t) =  a*e^{(b*t)}
\end{equation}

where  $a=  4.649*10^{-7}$ and $b=-0.004055$.

\begin{figure}[!ht]
\centering
\subfigure[]{
\includegraphics[width=0.45\textwidth]{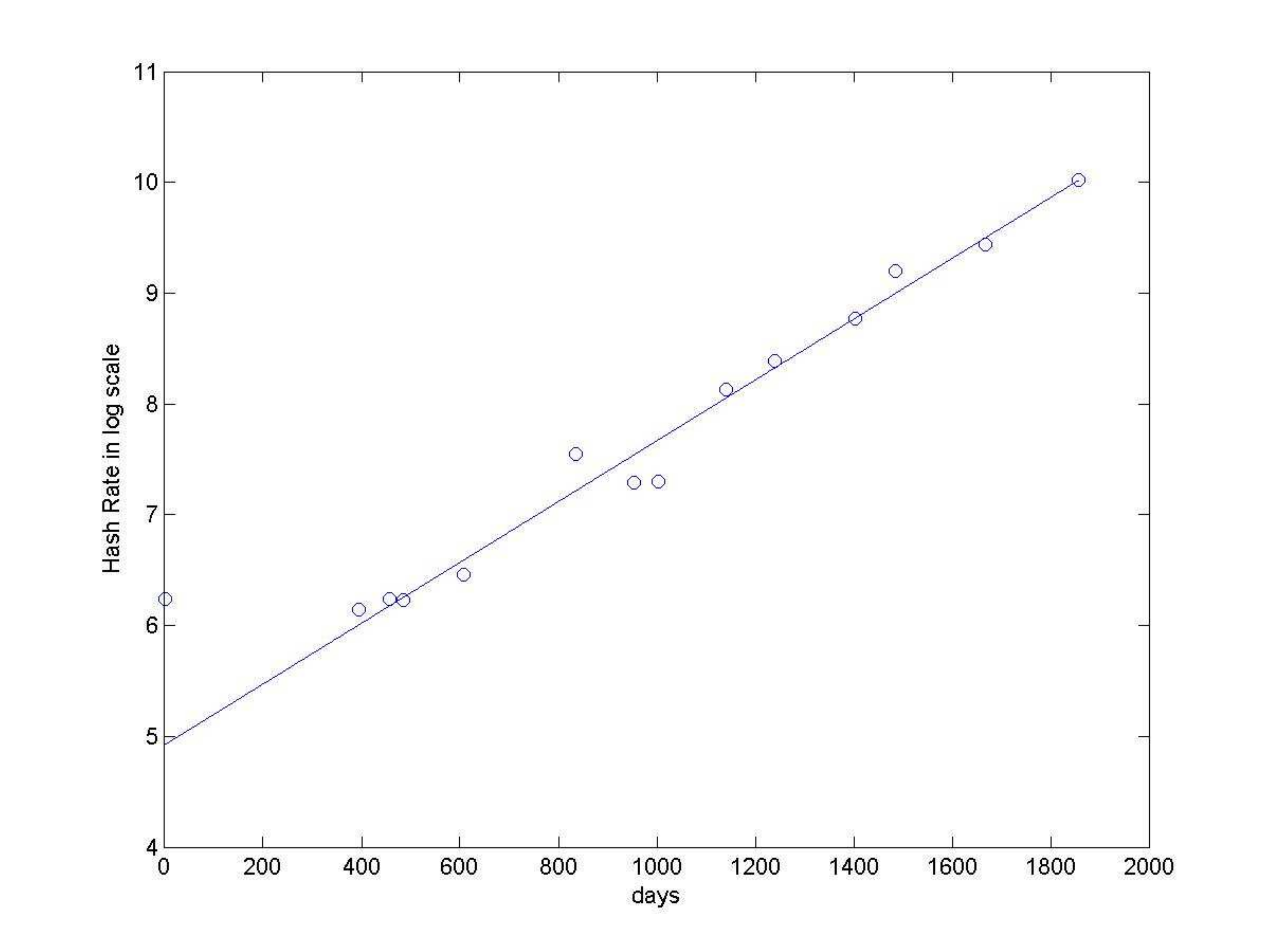}}
\hspace{7mm}
\subfigure[]{
\includegraphics[width=0.45\textwidth]{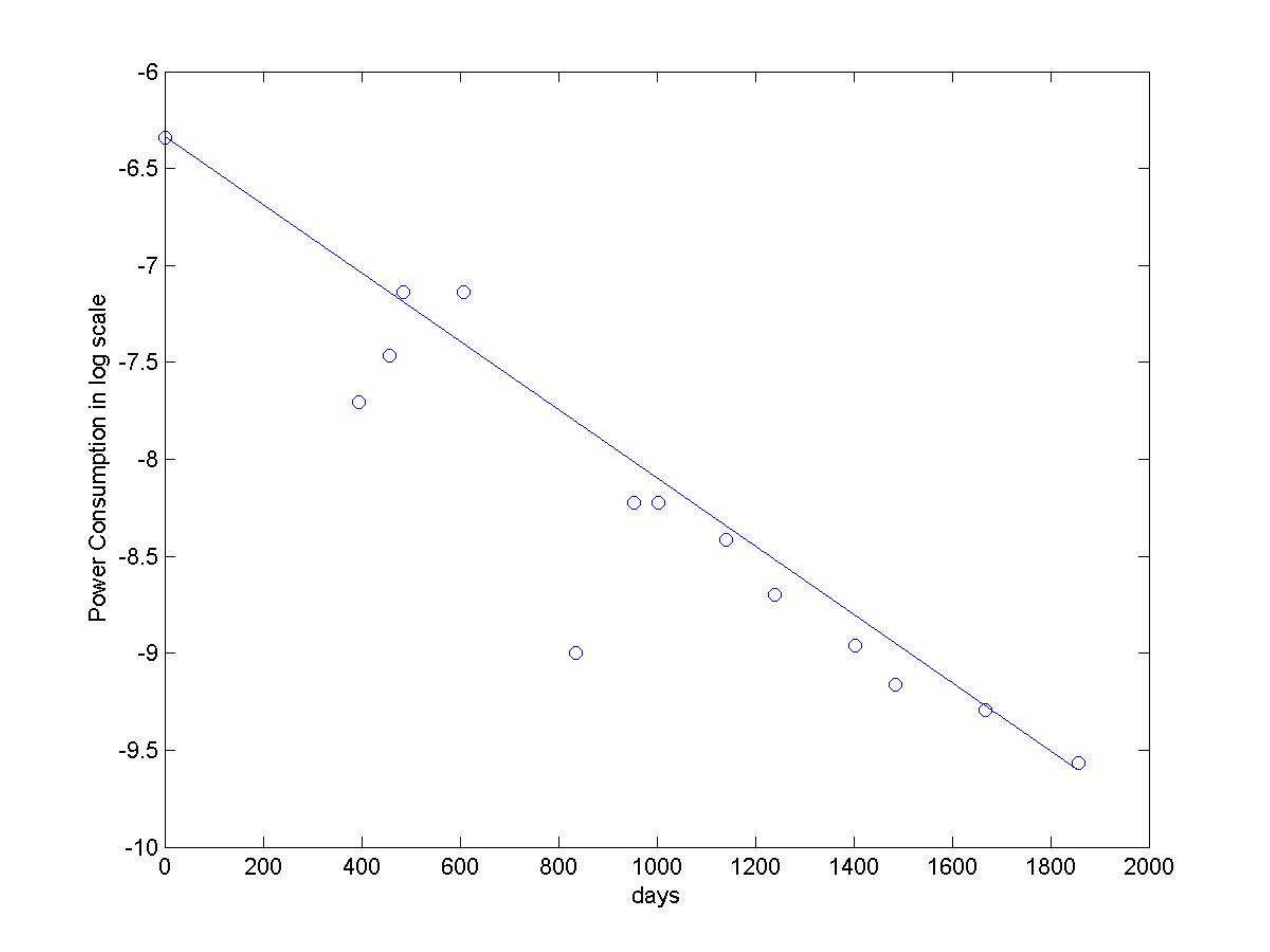}}
\caption{(a) Fitting curve of R(t) and (b) fitting curve of P(t). \label{fig:1}}
\end{figure}

Fig. \ref{fig:1} (a) and (b) show in logaritmic scale the fitting curves and how the hash rate increases over time, whereas power consumtpion decreases.

\section{The Model }\label{sec:4}

We used the \textit{Blockchain.info}, a web site which displays detailed information about all transactions and Bitcoin blocks, providing graphs and statistics on different data, for extracting the empirical data used in this work. This web site provides several graphs and statistical analysis of data about Bitcoins. 
In particular, we can observe the time trend of the Bitcoin price in the market, the total number of Bitcoins, the total hash rate of the Bitcoin network and the total number of Bitcoin transactions.

The proposed model presents an agent-based artificial cryptocurrency market in which agents mine, buy or sell Bitcoins. 

We modeled the Bitcoin market starting from September 1st, 2010, because one of our goals is to study the economy of the mining process. It was only around this date that miners started to buy mining hardware to mine Bitcoins, denoting a business interest in mining. Previously, they typically just used the power available on their personal computers.

The features of the model are:
\begin{itemize}
\item there are various kinds of agents active on the BTC market: Miners, Random traders and Chartists;
 \item the trading mechanism is based on a realistic order book that keeps sorted lists of buy and sell orders, and matches them allowing to fulfill compatible orders and to set the price;
 \item agents have typically limited financial resources, initially distributed following a power law;
 \item the number of agents engaged in trading at each moment is a fraction of the total number of agents;
 \item a number of new traders, endowed only with cash, enter the market; they represent people who decided to start trading or mining Bitcoins;
\item miners belong to mining pools. This means that at each time $t$ they always have a positive probability to mine at least a fraction of Bitcoin. Indeed, since 2010 miners have been pooling together to share resources in order to be able avoiding effort duplication to optimally mine Bitcoins. A consequence of this fact is that gains are smoothly distributed among miners.

On July 18th, 2010,
\begin{quote}
 \small{"ArtForz establishes an OpenGL GPU hash farm and generates his first Bitcoin block"} 
\end{quote}
 and on September 18th, 2010, 
\begin{quote}
\small{"Bitcoin Pooled Mining (operated by slush), a method by which several users work collectively to mine Bitcoins and share in the benefits, mines its first block"},
\end{quote}
(news from the web site http://historyofBitcoin.org/).
 \end{itemize} 

Since then, the difficulty of the problem of mining increased exponentially, and nowaday it would be almost unthinkable to mine without participating to a pool.

In the next subsections we describe in detail the model simulating the mining, the Bitcoin market and the related mechanism of Bitcoin price formation.

\subsection{The Agents}\label{sec:4.1}
Agents, or traders are divided into three populations: Miners, Random traders 
and Chartists.

Every \textit{i}-th trader enters the market at a given time step, $t_i^E$. Such a trader can be either a Miner, a Random trader or a Chartist.
All traders present in the market at the initial time $t_i^E=0$ holds an amount $c_i(0)$ of fiat currency (cash, in US dollars) and an amount $b_i(0)$ of crypto currency (Bitcoins), where $i$ is the trader's index. They represent the persons present in the market, mining and trading Bitcoins, before the period considered in the simulation.
Each \textit{i}-th trader entering the market at $t_i^E>0$ holds only an amount $c_i(t_i^E)$ of fiat currency (cash, in dollars). These traders represent people interested in entering the market, investing their money in it.

The wealth distribution of traders follows a Zipf law \cite{Levy}. 
The set of all traders entering the market at time $t_i^E>0$ are generated before the beginning of the simulation with a Pareto distribution of fiat cash, and then are randomly extracted from the set, when a given number of them must enter the market at a given time step.
Also, the wealth distribution in crypto cash of the traders in the market at initial time follows a Zipf law. Indeed, the wealth share in the world of Bitcoin is even more unevenly distributed than in the world at large (see web site http://www.cryptocoinsnews.com/owns-Bitcoins-infographic-wealth-distribution/).
More details on the trader wealth endowment are illustrated in Appendix.

\paragraph{Miners}\label{sec:4.1.1}

\textit{Miners} are in the Bitcoin market aiming to generate wealth by gaining Bitcoins.
At the initial time, the simulated Bitcoin network is calibrated respecting the Satoshi original idea of Bitcoin network where each node participates equally to the process of check and validation of the transactions and mining. We assumed that miners in the market at initial time ($t_i^E=0$) own a personal PC such as Core i5 2600K, and hence they are initially endowed with a hashing capability $r_i(0)$ equal to 0.0173GH/sec, that implies a power consumption equal to 75W \cite{Courtois}. 
Core i5 is a brand name of a series of fourth-generation x64 microprocessor developed by Intel and brought to market in October 2009.

Miners entering the market at time $t_i^E>0$ acquire mining hardware, and hence a hashing capability $r_i(t)$, which implies a specific electricity cost $e_i(t)$, investing a fraction $\gamma_{1,i}(t)$ of their fiat cash $c_i(t)$. 

In addition, over time all miners can improve their hashing capability by buying new mining hardware investing both their fiat and crypto cash. Consequently, the total hashing capability of $i-th$ trader at time $t$, $r_{i}(t)$ expressed in $[H/sec]$, and the total electricity cost $e_{i}(t)$ expressed in \$ per day, associated to her mining hardware units, are defined respectively as:

\begin{equation}
r_{i}(t) = \sum_{s=t_i^E}^{t}r_{i,u}(t)
\end{equation}

 and

\begin{equation}
e_i(t)=  \sum_{s=t_i^E}^{t}  \epsilon *P(s)*r_{i,u}(s)*24 
\end{equation}

where:
\begin{equation}
r_{i,u}(t=t_i^E>0) =  \gamma_{1,i}(t) c_i(t) R(t)
\end{equation}

\begin{equation}
r_{i,u}(t>t_i^E) =  [\gamma_{1,i}(t) c_i(t)+ \gamma_i(t) b_i(t) p(t)]R(t)
\end{equation}

\begin{itemize} 
\item $R(t)$ and $P(t)$ are, respectively, the hash rate which can be bought with one US\$, expressed in $\frac{H}{sec*\$}$, and the power consumption, expressed in $\frac{W}{H/sec}$. At each time $t$, their values are given by using the fitting curves described in subsection \textit{Modeling the Mining Hardware Performances};
\item $r_{i,u}(t)$ is the hashing capability of the hardware units $u$ bought at time $t$ by $i-th$ miner;
\item $\gamma_i=0$ and $\gamma_{1,i}=0$ if no hardware is bought by $i-th$ trader at time $t$. When a trader decides to buy new hardware, $\gamma_{1,i}$ represents the percentage of miner's cash devoted to buy it. It is equal to a random variable characterized by a lognormal distribution with average 0.15 and standard deviation 0.15.
$\gamma_{i}$ represents the percentage of miner's Bitcoins to be sold for buying the new hardware. It is equal to a random variable characterized by a lognormal distribution with average 0.175 and standard deviation 0.075. The term $\gamma_{1,i}(t) c_i(t)+ \gamma_i(t) b_i(t) p(t)$ expresses the amount of personal wealth that the miner wishes to devote to buy new mining hardware, meaning that on average the miner will devote 35\% of her cash and 17.5\% of her bitcoins to this purpose.
If $\gamma_i>1$ or $\gamma_{1,i}>1$, they are set equal to one;
\item $\epsilon$ is the fiat price per Watt and per hour. It is assumed equal to $1.4 \ast 10^{-4}$ \$, considering the cost of 1 KWh equal to 0.14\$,
that we assumed to be constant throughout the simulation. This electricity price is computed making an average of the electricity prices in the countries in which the Bitcoin nodes distribution is higher; see web sites \textit{https://getaddr.bitnodes.io} and $http://en.wikipedia.org/wiki/Electricity\_pricing$.
\end{itemize}

The decision to buy or not new hardware is taken by every miner from time to time, on average every two months (60 days). If $i-th$ miner decides whether to buy new hardware and/or to divest the old hardware units at time $t$, the next time, $t^{I-D}_i(t)$, she will decide again is given by eq. \ref{ID}:

\begin{equation}\label{ID}
t^{I-D}_i(t)=t+ int(60+ N(\mu^{id},\sigma^{id}))
\end{equation}

where $int$ rounds to the nearest integer and $N(\mu^{id},\sigma^{id})$ is a normal distribution with average $\mu^{id}=0$ and standard deviation $\sigma^{id}= 6$.
$t^{I-D}_i(t)$ is updated each time the miner takes her decision.

Miners active in the simulation since the beginning will take their first decision within 60 days, at random times uniformly distributed. Miners entering the simulation at time $t>1$ will immediately take this decision.

In deeper detail, at time $t=t^{I-D}_i(t)$, every miner can decide to buy new hardware units, if her fiat cash is positive, and/or to divest the old hardware units. If trader's cash is zero, she issues a sell market order to get the cash to support her electricity expenses, $c_{i,a}(t)=\gamma_i(t) b_i(t) p(t)$.

Each $i-th$ miner belongs to a pool, and consequently at each time $t$ she always has a probability higher than 0 to mine at least some sub-units of Bitcoin. This probability is inversely proportional to the hashing capability of the whole network. Knowing the number of blocks discovered per day, and consequently knowing the number of new Bitcoins $B$ to be mined per day, the number of Bitcoins $b_{i}$ mined by $i-th$ miner per day can be defined as follows: 

\begin{equation}
b_{i}(t)= \frac{r_{i}(t) }{ r_{Tot}(t)}B(t)
\end{equation}

where: 
\begin{itemize}
\item $r_{Tot}(t)$  is the hashing capability of the whole population of miners $N_m$ at time $t$ defined as the sum of the hashing capabilities of all miners at time $t$, $ \sum_i^{N_m} r_{i}(t)$;
\item the ratio $\frac{r_{i}(t) }{ r_{Tot}(t)}$ defines the relative hash rate of $i-th$ miner at time $t$.
\end{itemize}

Note that, as already described in section \textit{Mining Process}, the parameter $B$ decreases over time. At first, each generated block corresponds to the creation of 50 Bitcoins, but after four years, such number is halved. So, until November 27, 2012, 100,800 Bitcoins were mined  in 14 days (7200 Bitcoins per day), and then 50,400 Bitcoins in 14 days (3600 per day).

The decision of a miner to buy and/or divests a hardware unit $u$ depends on the Bitcoins potentially obtained mining with the new hardware. 
A miner buys new hardware units if the daily cost, given by the expense in electricity, $e_{i,u}(t)$, associated to these units is smaller than the gain expected in Bitcoin.  
Hence, before buying new hardware units the following constraint has to be evaluated:

\begin{equation}\label{ConstraintToBuy}
e_{i,u}(t) < b_{i,u}(t) p(t)
\end{equation}

where:
\begin{itemize}
\item $b_{i,u}$ are the Bitcoins potentially mined by unit $u$ at time $t$: $b_{i,u}(t)= \frac{r_{i,u}(t) }{ r_{Tot}(t)}B(t)$
\item $p(t)$ is the Bitcoin price at time $t$.
\end{itemize}

Only if this constraint is respected the miner can buy new hardware. In this case, she issues a market order acquiring an amount of fiat cash $c_{i,a}(t)=\gamma_i(t) b_i(t) p(t)$ in the next time steps. She invests 50\% of this amount to buy new hardware and keeps the remaining 50\% as cash, to pay the electricity bill for her hardware.

If the constraint in eq. \ref{ConstraintToBuy} is not respected, the miner anyway issues a market order equal to $c_{i,a}(t)=\frac{\gamma_i(t) b_i(t) p(t)}{2}$ to support her electricity expenses.

A miner divests her old hardware units if the expense in electricity associated to that units is 20\% higher than  the gain expected in Bitcoins using that hardware, at the current price. 
Therefore, the following constraint has to be respected for each value of $k$, with $k$ going from 0 to current time $t$:

\begin{equation}\label{ConstraintToDivest}
e_{i,u}(k) \le 1.2 \frac{r_{i,u}(k) }{ r_{Tot}(t)} B(t) p(t)
\end{equation}

The model also includes a mechanism that enables 10\% of miners to invest and/or divest their hardware also at a time $t \ne t^{I-D}_i(t)$. This mechanism is triggered when the price relative variation, in a time window $\tau^M$ equal to 15 days, is positive and is higher than a threshold $Th^M$ equal to 0.016.
This because, in the real market, the investments of miners grow when the profitability of mining activity increases. So, increasing the interest of miners in buying new hardware in these periods is a plausible assumption.

\paragraph{Random Traders}\label{sec:4.1.2}
\textit{Random Traders} represent persons who enter the crypto-currency market for various reasons, but not for speculative purposes. They issue orders for reasons linked to their needs, for instance they invest in Bitcoins to diversify their portfolio, or they disinvest to satisfy a need for cash. They issue orders in a random way, compatibly with their available resources. In particular, buy and sell orders are always issued with the same probability. The specifics of their behavior is described in section \textit{Buy and Sell Orders}.
 
\paragraph{Chartists}
\textit{Chartists} represent speculators, aimed to gain by placing orders in the Bitcoin market. 
They speculates that, if prices are rising, they will keep rising, and if prices are falling, they will keep falling. 
In particular, $i-th$ chartist issues a buy order when the price relative variation in a time window $\tau_i^C$, is higher than a threshold $Th^C=0.01$, and issues a sell order if this variation is lower than $Th^C$. 
$\tau_i^C$ is specific for each chartist, and is characterized by a normal distribution with average equal to 20 and standard deviation equal to 1.
Chartists usually issue buy orders when the price is increasing and sell orders when the price is decreasing. However, 10\% of Chartists decide, instead, to adopt a contrary strategy, and place a sell order instead of a buy order, or vice-versa. This contrarian behavior is common in financial markets, and is typically modeled also in market models \cite{Raberto2003}.
Note that a Chartist will issue an order only when the price variation is above a given threshold. So, in practice, the extent of Chartist activity varies over time.
In general the modelled Chartists' behavior is key to produce large price variations, and to the reproduction of the basic statistical proprieties of the real returns.

All Random traders and Chartists entering the market at $t=t^E > 0$, issue a buy order to acquire their initial Bitcoins. Over time, at time $t>t^E$ only a fraction of Random traders and Chartists is active, and hence enabled to issue orders. 
Active traders can issue only one order per time step, which can be a sell order or a buy order. 

Orders already placed but not yet satisfied or withdrawn are accounted for when determining the amount of Bitcoins a trader can buy or sell. 
Details on the percentage of active traders, the number of the traders in the market and on the probability of each trader to belong to a specific traders' population are described in Appendix.

\subsection{Buy and Sell Orders}\label{sec:4.2}
The Bitcoin market is modelled as a steady inflow of buy and sell orders, placed by the traders as described in \cite{Cocco2014}.
Both buy and sell orders are expressed in Bitcoins, that is, they refer to a given amount of Bitcoins to buy or sell.
In deeper detail, all orders have the following features:
\begin{itemize}
\item amount, expressed in \$ for buy order and in Bitcoins for sell order: the latter amount is a real number, because Bitcoins can be bought and sold in fractions as small as a "Satoshi";
\item residual amount (Bitcoins or \$): used when an order is only partially satisfied by previous transactions;
\item limit price (see below), which in turn can be a real number;
\item time when the order was issued;
\item expiration time: if the order is not (fully) satisfied, it is removed from the book at this time.
\end{itemize}
		
The amount of each buy order depends on the amount of cash, $c_i(t)$, owned by $i$-th trader at time $t$, less the cash already committed to other pending buy orders still in the book. Let us call $c^b_i$ the available cash. The number of Bitcoins to buy, $b_a$ is given by eq. \ref{eq-buy}

\begin{equation}\label{eq-buy}
b_a = \frac{c^b_i \beta}{ p(t)} 
\end{equation}

where $p(t)$ is the current price and $\beta$ is a random variable drawn from a lognormal distribution with average and standard deviation equal to $0.25$ and $0.2$, respectively for Random traders and equal to $0.4$ and $0.2$, respectively for Chartists. In the unlikely case that $\beta > 1$, $\beta$ is set equal to 1. 

Similarly, the amount of each sell order depends on the number of Bitcoins, $b_i(t)$ owned by $i$-th trader at time $t$, less the Bitcoins already committed to other pending sell orders still in the book, overall called $b^s_i$. 
The number of Bitcoins to sell, $s_a$ is given by 

\begin{equation}\label{eq-sell}
s_a = b^s_i \beta
\end{equation}

where $\beta$ is a lognormal random variable as above.
Short selling is not allowed. 

The limit price models the price to which a trader desire to conclude his/her transaction. 
An order can also be issued with no limit
(market order), meaning that its originator wishes to perform the trade at the best price she can find. 
In this case, the limit price is set to zero. 
The probability of placing a market order, $P_{lim}$, is set at the beginning of the simulation 
and is equal to 1 for Miners, to 0.2 for Random Traders and to 0.7 for Chartists.
This because, unlike Random Traders, if Miners and Chartists issue orders, then they wish to perform the trade at the best available price, the formers because they need cash, the latters to be able to gain by following the price trend.

Let us suppose that $i$-th trader issues a limit order to buy $a_i^b(t)$ Bitcoins at time $t$. 
Each buy order can be executed if the trading price is lower than, or equal to, its buy limit price $b_{i}$. 
In the case of a sell order of $a_i^s(t)$ Bitcoins, it can be executed if the trading price is higher than, or equal to, its sell limit price $s_{i}$.
As said above, if the limit prices $b_{i}=0$ or $s_{i}=0$, then the orders can be always executed, provided there is a pending complementary order.

The buy and sell limit prices, $b_{i}$ and $s_{i}$, are given respectively by the following equations:

\begin{equation}\label{buyLimit}
b_{i}(t)=p(t)*N_i(\mu,\sigma_i)
\end{equation}

\begin{equation}\label{sellLimit}
s_{i}(t)=\frac{p(t)}{N_i(\mu,\sigma_i)}
\end{equation}

where
\begin{itemize}
\item $p(t)$ is the current Bitcoin price; 
\item $N_i(\mu,\sigma^{c}_i)$ is a random draw from a Gaussian distribution with average $\mu \simeq 1$ and standard deviation $\sigma_i \ll 1$.
\end{itemize}

The limit prices have a random component, modelling the different perception of Bitcoin value, that is the fact that what traders "feel" is the right price to buy or to sell is not constant, and may vary for each single order. 
In the case of buy orders, we stipulate that a trader wishing to buy must offer a price that is, on average, slightly higher than the market price. 

The value of $\sigma_i$ is proportional to the "volatility"  $\sigma(T_i)$ of the price $p(t)$ through the equation $\sigma_i=K\sigma(T_i)$, where $K$ is a constant and $\sigma(T_i)$ is the standard deviation of price absolute returns, calculated in the time window $T_i$.
$\sigma_i$ is constrained between a minimum value $\sigma_{min}$ and a maximum value $\sigma_{max}$ (this is an approach similar to that of \cite{Raberto2001}). 
For buy orders $\mu=1.05$, $K=2.5$, $\sigma_{min}=0.01$ and $\sigma_{max}=0.003$.

In the case of sell orders, the reasoning is dual. For symmetry, the limit price is divided by a random draw from the same Gaussian distribution $N_i(\mu,\sigma^{c}_i)$.

An expiration time is associated to each order. 
For Random Traders, the value of the expiration time is equal to the current time plus a number of days (time steps) drawn from a lognormal distribution with average and standard deviation equal to 3 and 1 days, respectively. In this way, most orders will expire within 4 days since they were posted.
Chartists, who act in a more dynamic way to follow the market trend, post orders whose expiration time is at the end of the same trading day.
Miners issue market orders, so the value of the expiration time is set to infinite.

\subsection{Price Clearing Mechanism}\label{sec:4.3}

We implement the price clearing mechanism by using an Order Book similar to that presented in \cite{Raberto2005}.

At every time step, the order book holds the list of all the orders received and
still to be executed. 
Buy orders are sorted in descending order with respect to the limit price $b_{i}$. Orders with the same limit price are sorted in ascending order with respect to the order issue time.
Sell orders are sorted in ascending order with respect to the limit price $s_{j}$. Orders with the same limit price are sorted in ascending order with respect to the order issue time.

At each simulation step, various new orders are inserted into the respective lists. 
As soon as a new order enters the book, the first buy order and the first sell order of the lists are inspected to verify if they match. 
If they match, a transaction occurs. The order with the smaller residual amount is fully executed, 
whereas the order with larger amount is only partially executed, and remains in the head of the list, with its residual amount reduced by the amount of the matching order. Clearly, if both orders have the same residual amount, they are both fully executed.

After the transaction, the next pair of orders at the head of the lists are checked for matching. If they match, they are executed, and so on until they do not match anymore. 
Hence, before the book can accept new orders, all the matching orders are satisfied.

A sell order of index $j$ matches a buy order of index $i$, and vice versa, only if $s_{j} \le b_{i}$, or if one of the two limit prices, or both, are equal to zero. 

As regards the price, $p_T$, to which the transaction is performed, the price formation mechanism follows the rules described below. Here, $p(t)$ denotes the current price:
\begin{itemize}
\item when one of the two orders has limit price equal to zero: 
\begin{itemize}
\item if $b_{i} > 0$, then $p_T=min(b_{i}, p(t))$,
\item if $s_{j} > 0$, then $p_T=max(s_{j}, p(t))$,
\end{itemize}
\item when both orders have limit price equal to zero, $p_T = p(t)$;
\item  when both orders have limit price higher than zero, $p_T = \frac{b_{i}+s_{j}}{2}$.
\end{itemize}

\section{Simulation Results}\label{sec:6}
The model described in the previous section was implemented in Smalltalk language. Before the simulation, it had to be calibrated in order to reproduce the real stylized facts and the mining process in the Bitcoin market in the period between September 1st, 2010 and September 30, 2015. The simulation period was thus set to 1856 steps, a simulation step corresponding to one day. We included also weekends and holidays, because the Bitcoin market is, by its very nature, accessible and working everyday.

We set the initial value of several key parameters of the model by using data recovered from the Blockchain Web site. 
The main assumption we made is to size the artificial market at about 1/100 of the real market, to be able to manage the computational load of the simulation.
Table \ref{tab:initial} shows the parameter values and their computation assumptions in detail. 

In Appendix other details about the calibration of the model are shown. Specifically, the calibration of the trader wealth endowment, the number of active traders, the total number of traders in the market and the probability of a trader to belong to a specific traders' population are described in detail.

\begin{table}%
\caption{Values of simulation parameters and the assumptions behind them.
\label{tab:initial}}{%
\begin{tabular}{ccp{8cm}}
\hline\noalign{\smallskip}
Param.&Initial Value&Description and discussion\\
\noalign{\smallskip}\hline\noalign{\smallskip}
$N_t(0)$&160&Number of initial traders. Obtained dividing by 100 the number of traders on September 1st, 2010 estimated through the fitting curve shown in eq. \ref{N_T} (see Appendix).\\
\hline
$N_t(T)$&39,649&Total number of traders at the end of the simulation. Obtained dividing by 100 the number of traders on September 30, 2015 estimated through the fitting curve shown in eq. \ref{N_T}.\\
\hline
$B$&72 or 36&Bitcoins mined per day. Obtained dividing by 100 the Bitcoins which are mined every day. They are 72 until $853th$ simulation step (November 27th, 2012), and 36 from $853th$ simulation step onwards.\\
\hline
$p(0)$&0.0649 \$&Initial price. The average price as of September 2010.\\
\hline
$B_T(0)$& 23,274 \$&Total initial crypto cash. Obtained dividing by 100 the number of Bitcoins on September 1st, 2010 and keeping just 60\% of this value, because we assume that 40\% of Bitcoins are not available for the  trade.\\
\hline
$q$& 200,000 \$&Constant used in Zipf's law ($\frac{q}{i^{0.6}}$), used to assign the initial cash for traders entering at $t>1$.\\
\hline
$c^s_1$& 20,587 \$&Initial cash of the richest trader entering the simulation at $t=1$.\\
\hline
$b^s_1$& 4,117 \$&Initial Bitcoin cash of the richest trader entering the simulation at $t=1$.\\
\noalign{\smallskip}\hline
\end{tabular}}
\end{table}

The model was run to study the main features which characterize the Bitcoin market and the traders who operate in it.
In order to assess the robustness of our model and the validity of our statistical analysis, we
repeated 100 simulations with the same initial conditions, but different seeds of the random number generator. 
The results of all simulations were consistent, as shown in the followings.

\subsection{Bitcoin prices in the real and simulated market}\label{sec:6.1}

\begin{figure}
\centering
 \includegraphics[width=0.8\textwidth]{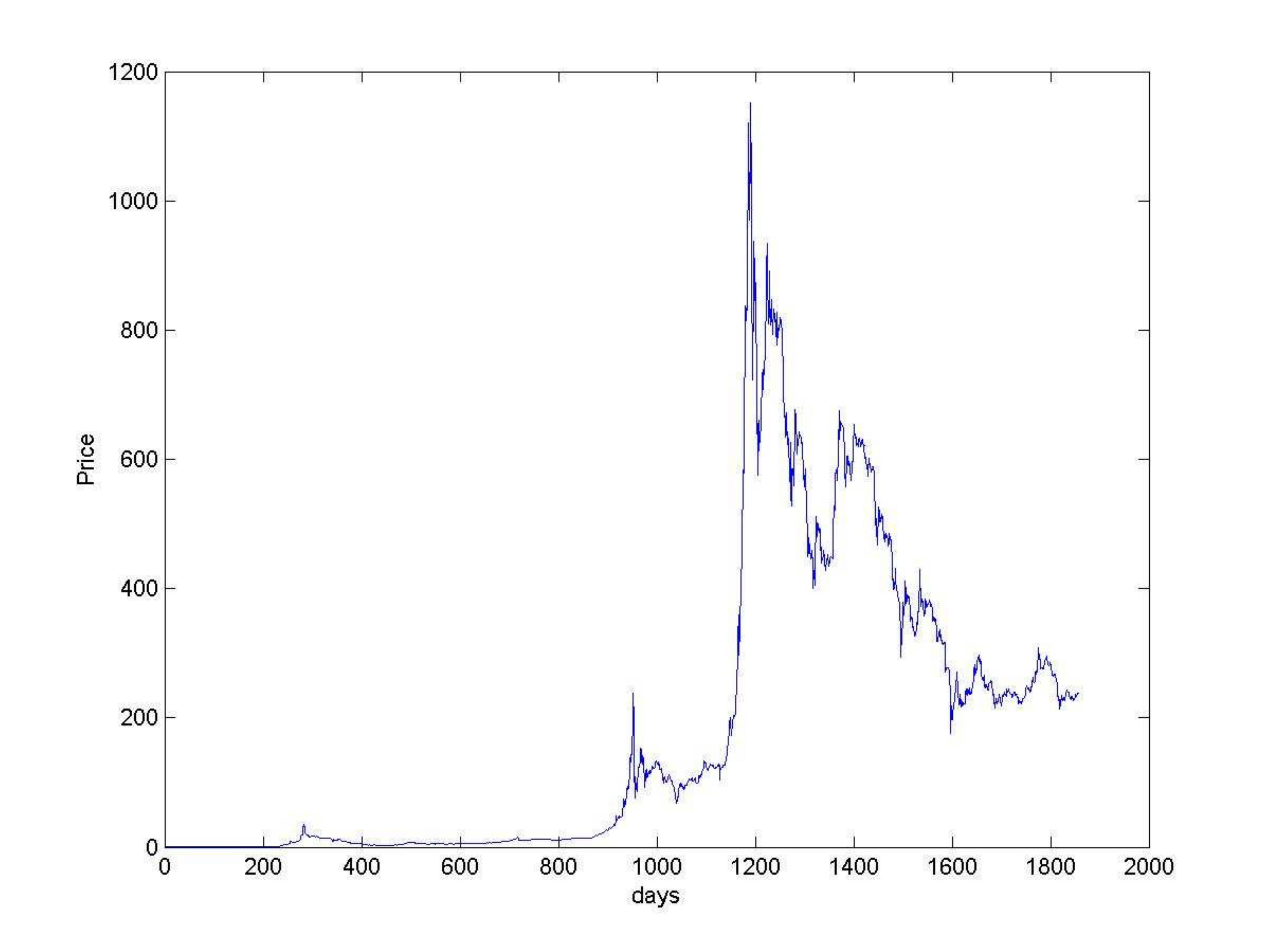}
\caption{Price of Bitcoins in US\$.}
\label{fig:realPrice}
\end{figure}

We started studying the real Bitcoin price series between September 1st, 2010 and September 30, 2015, shown in 
Fig. \ref{fig:realPrice}. The figure shows an initial period in which the price trend is relatively constant, until about $950^{th}$ day. Then, a period of volatility follows between $950^{th}$ and $1150^{th}$ day, followed by a period of strong volatility, until the end of the considered interval. The Bitcoin price started to fall at the beginning fo 2014, and is continuing on its downward slope until September 2015.

It is well known that the price series encountered in financial markets typically exhibit some
statistical features, also known as "stylized facts" \cite{Pagan,Lux}. Among these, the three 
uni-variate properties which appear to be the most important and pervasive of price series, are 
(i) the unit-root property, (ii) the fat tail phenomenon, and (iii) the \emph{Volatility Clustering}.
We examined daily Bitcoin prices and found that also these prices exhibit these properties as discussed in detail in \cite{Cocco2014}.

As regards the prices in the simulated market, we report in Fig. \ref{fig:Price} the Bitcoin price in one
typical simulation run. It is possible to observe that, as in the case of the real price, at first the price keeps its value constant, but then, after about 1000 simulation steps, contrary to what happens in the reality, it grows and continues on its upward slope until the end of the simulation period. 

\begin{figure}
\centering
 \includegraphics[width=0.5\textwidth]{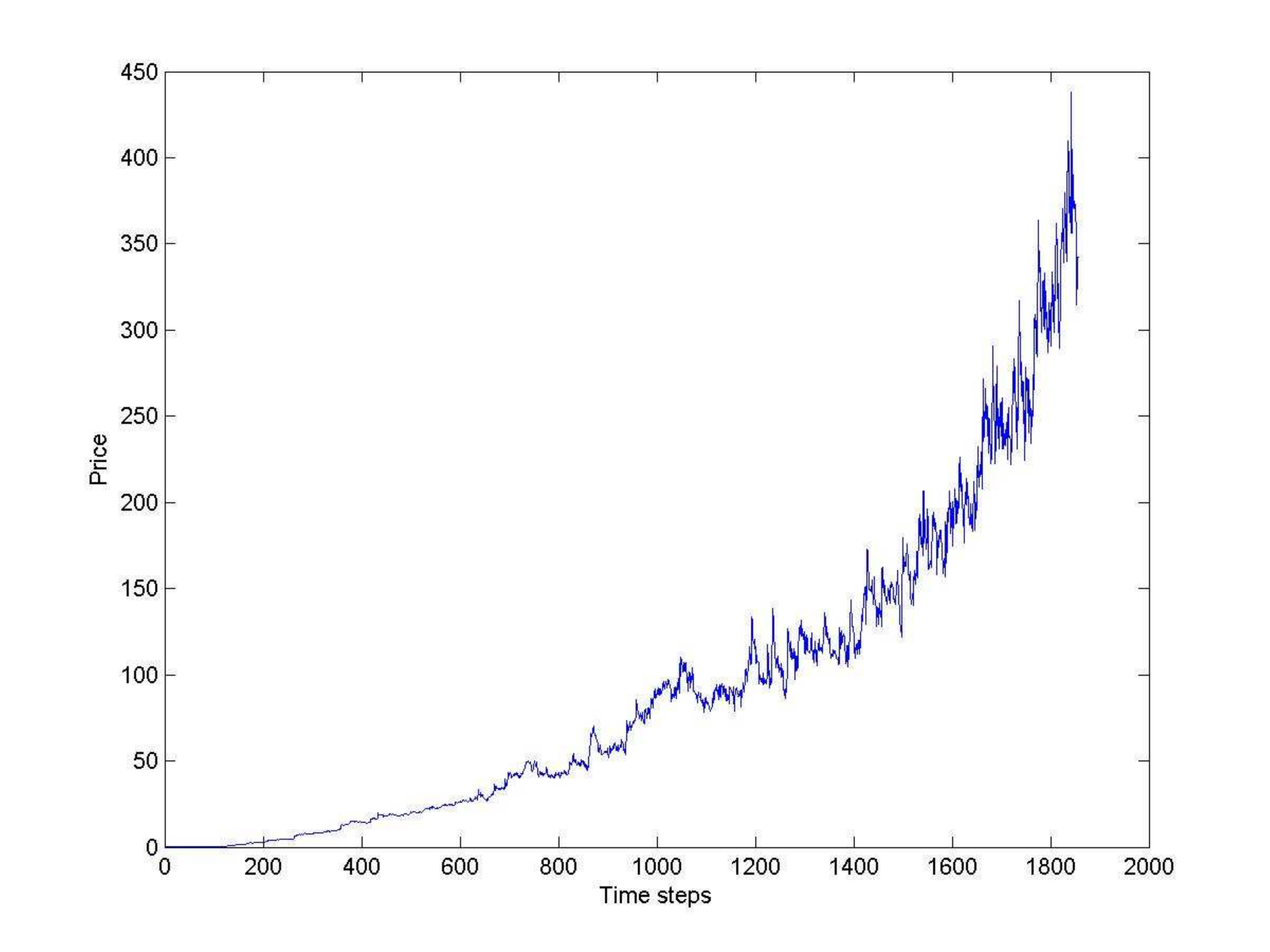}
\caption{Bitcoin simulated Price in one simulation run.}
\label{fig:Price}      
\end{figure}

Figs. \ref{fig:averagePrice} (a) and (b) report the average and the standard deviation of the simulated price, taken on all 100 simulations. Note that the average value of prices steadily increases with time, in contrast with what happens in reality.
Fig. \ref{fig:averagePrice} (b) shows that the price variations in different simulation runs increase with time, as the number of traders, transactions and the total wealth in the market are increasing.

\begin{figure}[!ht]
\centering
\subfigure[]{
\includegraphics[width=0.45\textwidth]{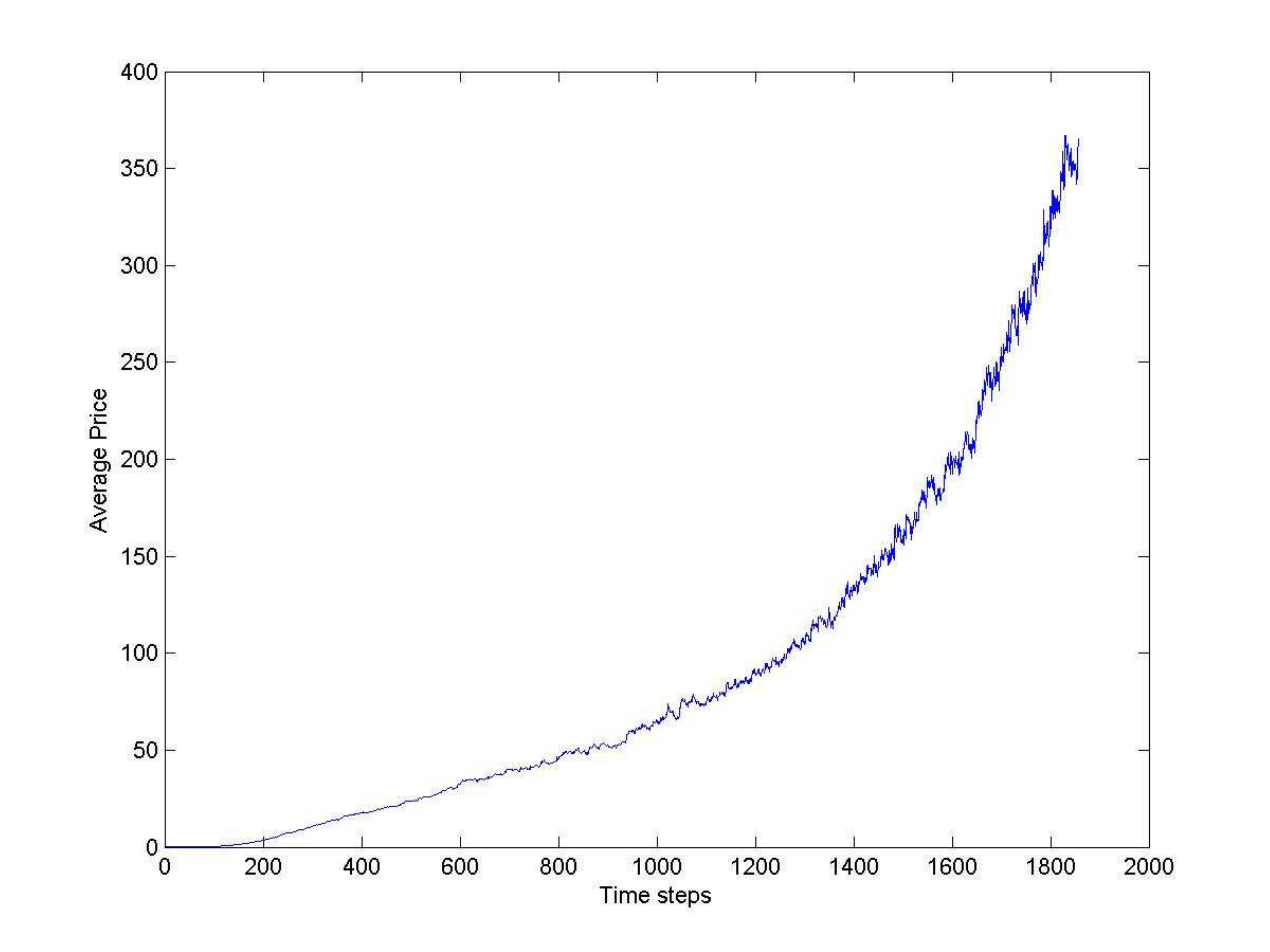}}
\hspace{7mm}
\subfigure[]{
\includegraphics[width=0.45\textwidth]{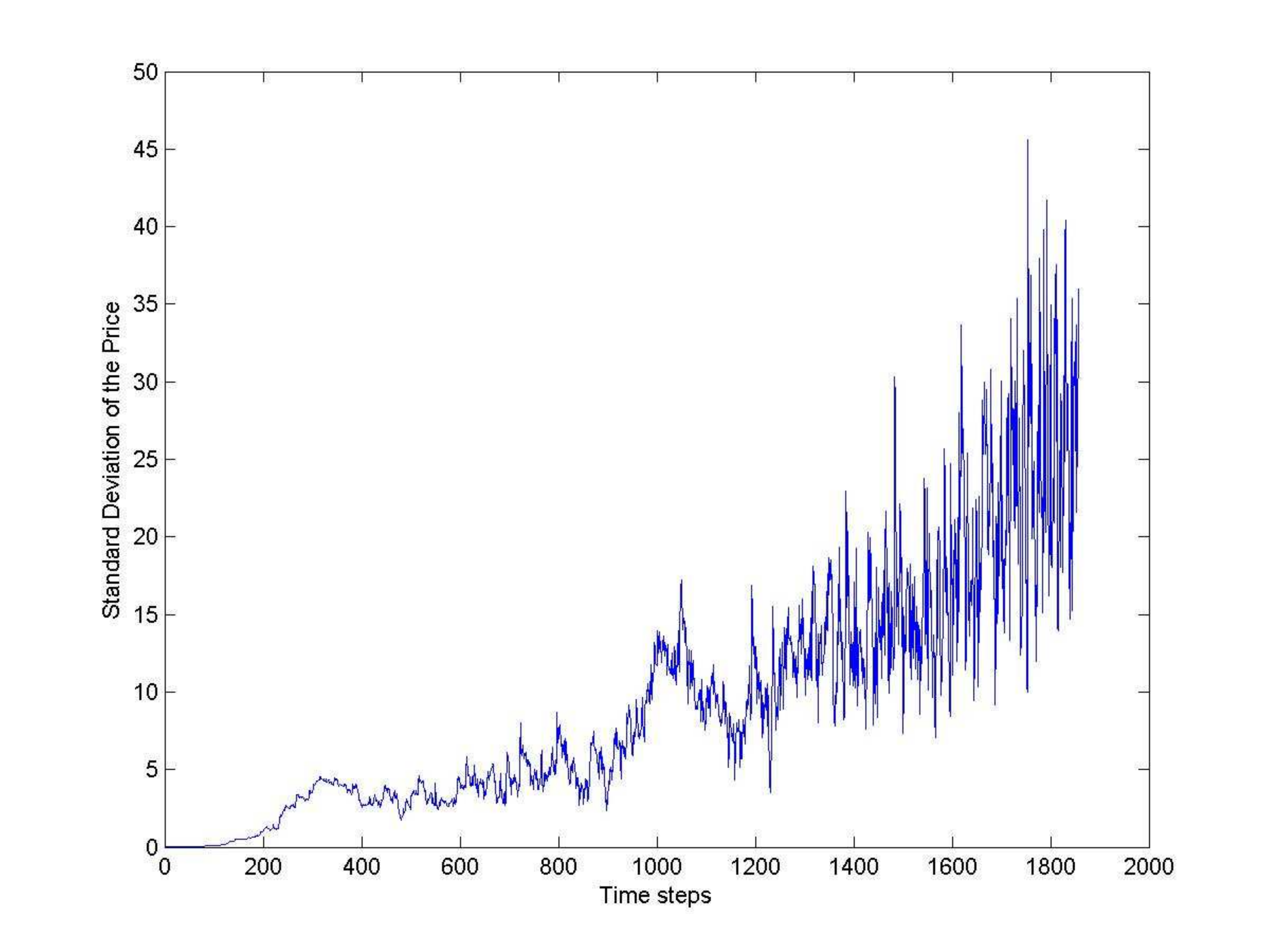}}
\caption{(a) Average Price  and (b) standard deviation computed on the 100 Monte Carlo simulations performed. \label{fig:averagePrice}}
\end{figure}

In the proposed model, the upward trend of the price depends on an intrinsic mechanism -- in fact, the average price tends to the ratio of total available cash to total available Bitcoins. Since new traders bring in more cash than new mined Bitcoins, the price tends to increase.

In reality, Bitcoin price is also heavily affected by exogenous factors. For instance, in the past the price strongly reacted to reports such as those regarding Bitcoin ban in China, or the MtGox exchange going bust. Moreover, the total capitalization of the Bitcoin market is of the order of just some billions of US\$, so if a large hedge fund decided to invest in Bitcoins, or if large amounts of Bitcoins disappeared because of theft, fraud or mismanagement, the effect on price would be potentially quite large. 
All these exogenous events, that can trigger strong and unexpected price variations, obviously cannot be part of our base model.

In section \textit{Other Results}, we shall describe the results obtained when some random traders adopt speculative behaviors, in addition to the speculative behaviour that characterizes Chartists. Simulating this behavior allows to reproduce the Bitcoin price peak in December 2013 and its subsequent fall. 

Despite inability to reproduce the decreasing trend of the price, the model presented in Section \textit{The Model}, is able to reproduce quite well all statistical properties of real Bitcoin prices and returns. The stylized facts, robustly replicated by the proposed model, are the same of a previous work of Cocco et al. \cite{Cocco2014}, and do not depend on the addition of the miners to the model.

\subsection{Traders' Statistics}\label{sec:6.3}

\begin{figure}[!ht]
\centering
\subfigure[]{
\includegraphics[width=0.45\textwidth]{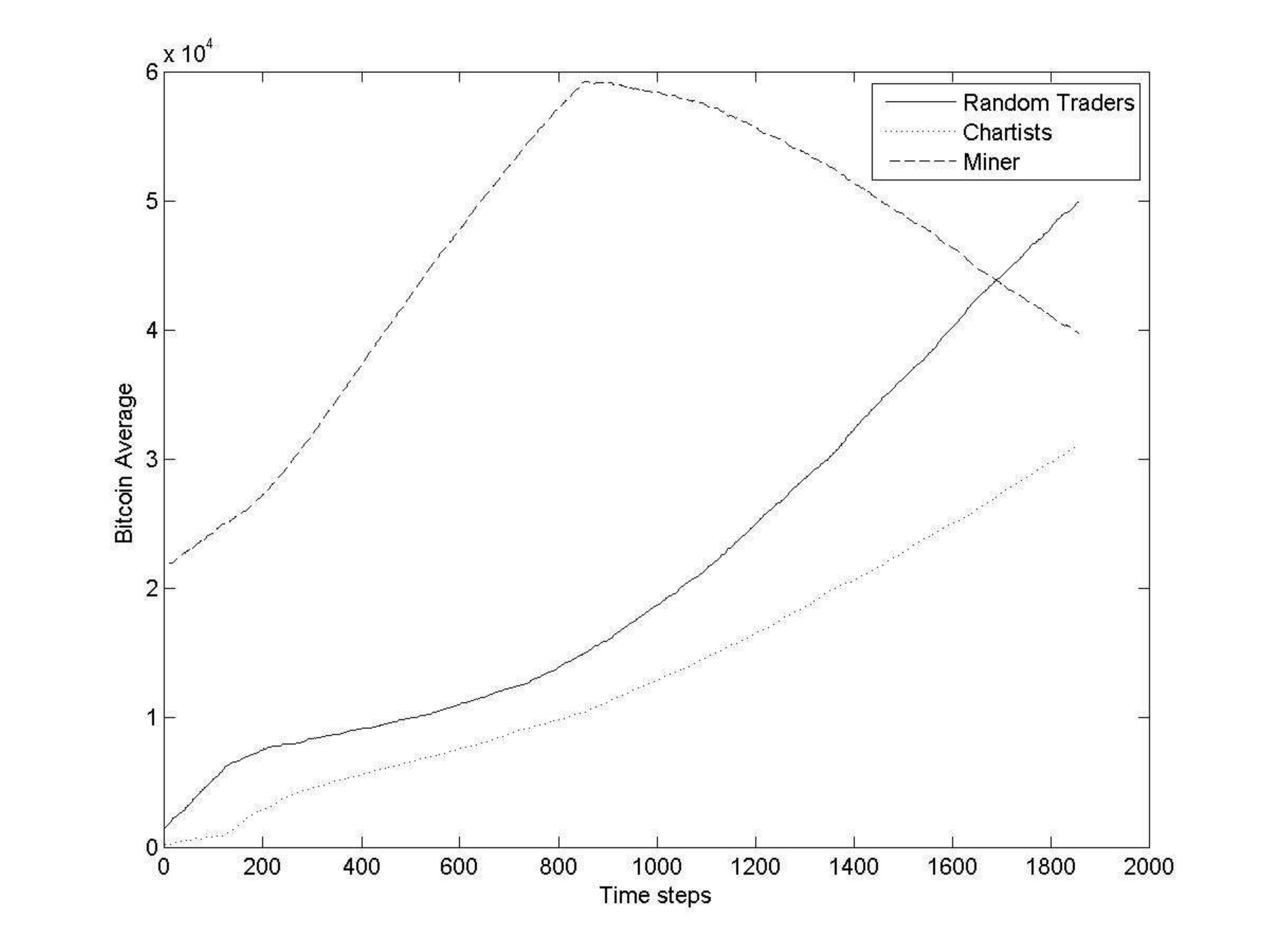}}
\hspace{4mm}
\subfigure[]{
\includegraphics[width=0.45\textwidth]{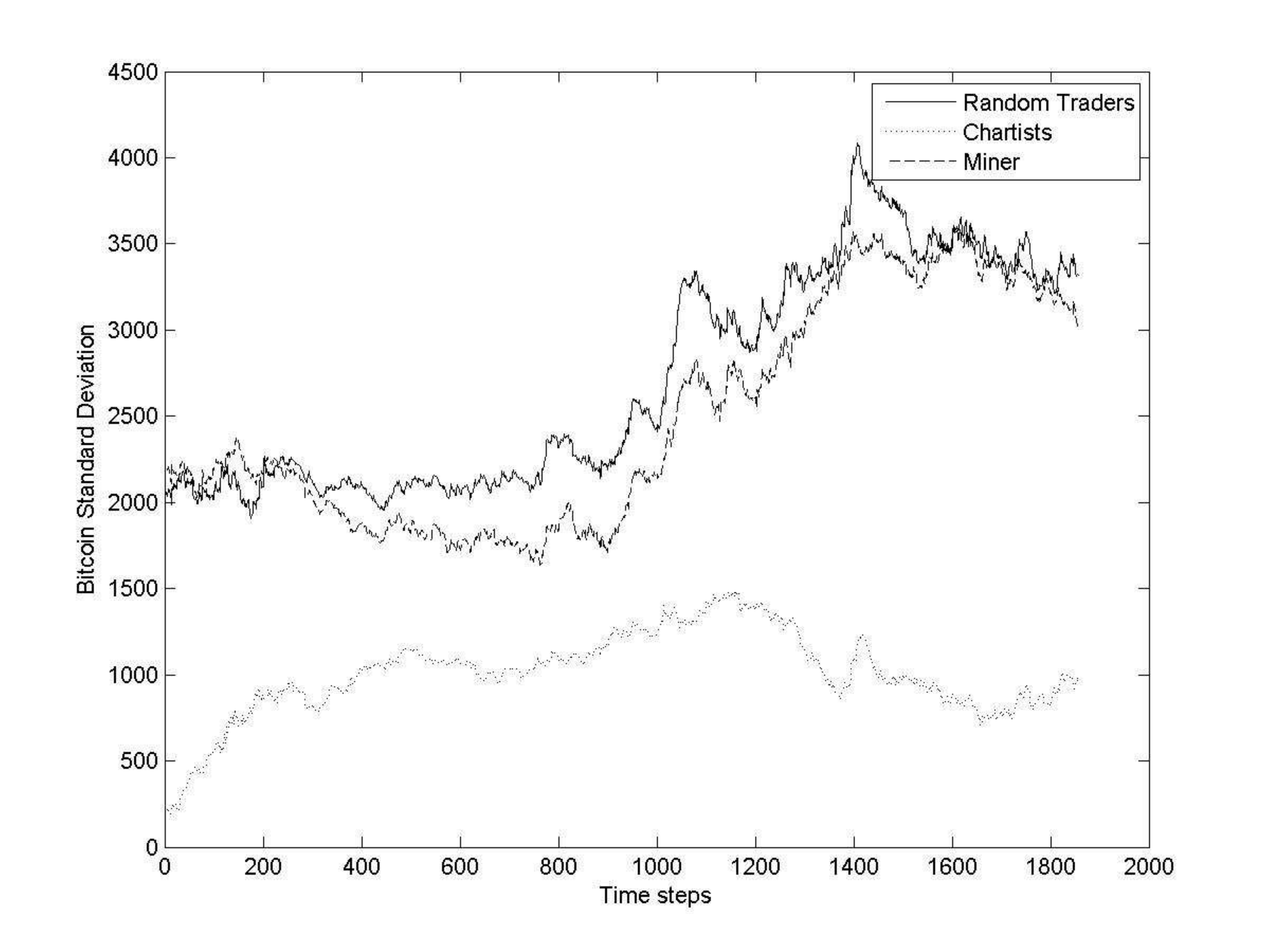}}
\caption{(a) Average and (b) standard deviation of the Bitcoin amount for all trader populations during the simulation period across all Monte Carlo simulations. \label{fig:AvgStdBtc}}
\end{figure}

\begin{figure}[!ht]
\centering
\subfigure[]{
\includegraphics[width=0.45\textwidth]{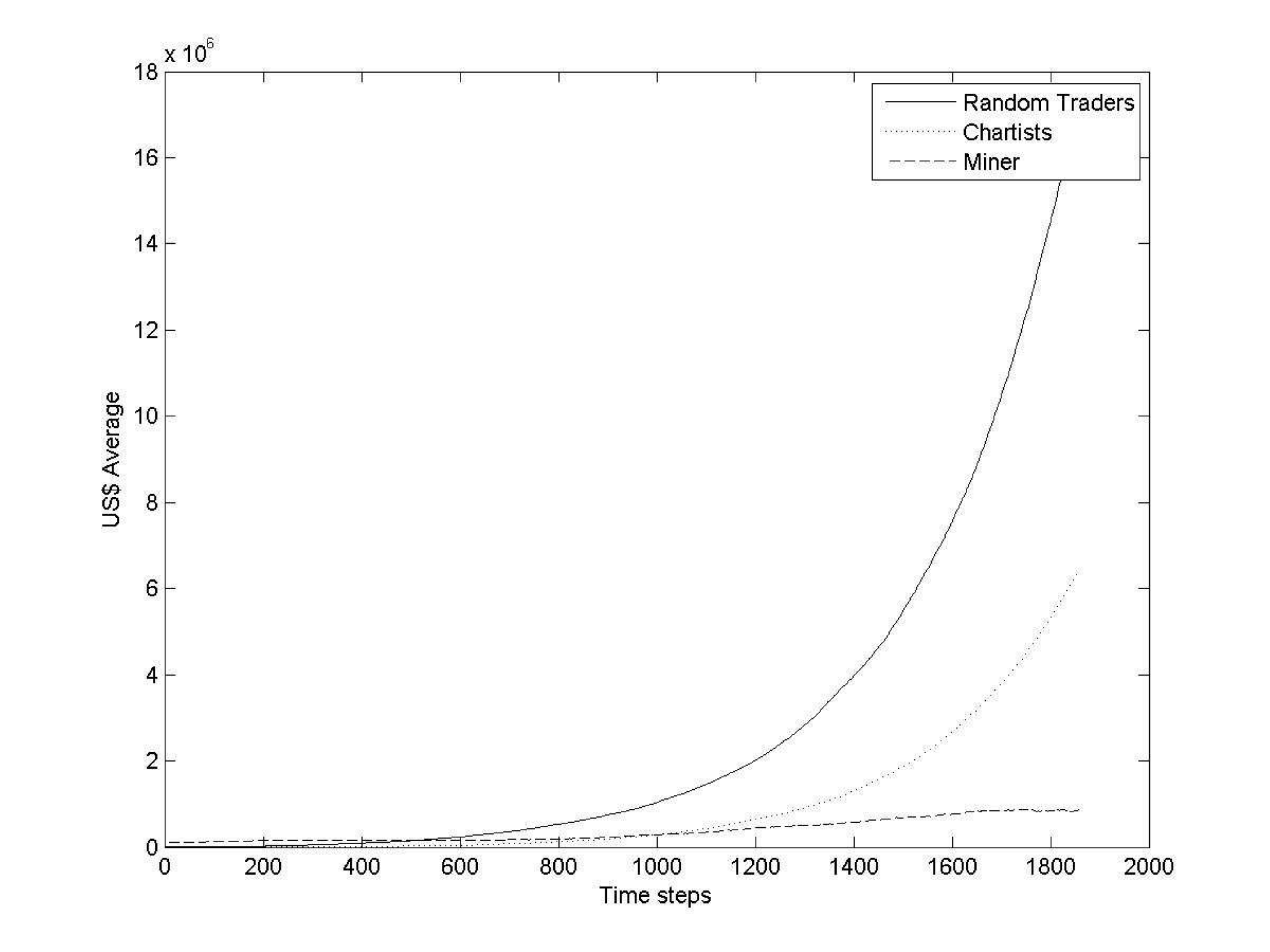}}
\hspace{4mm}
\subfigure[]{
\includegraphics[width=0.45\textwidth]{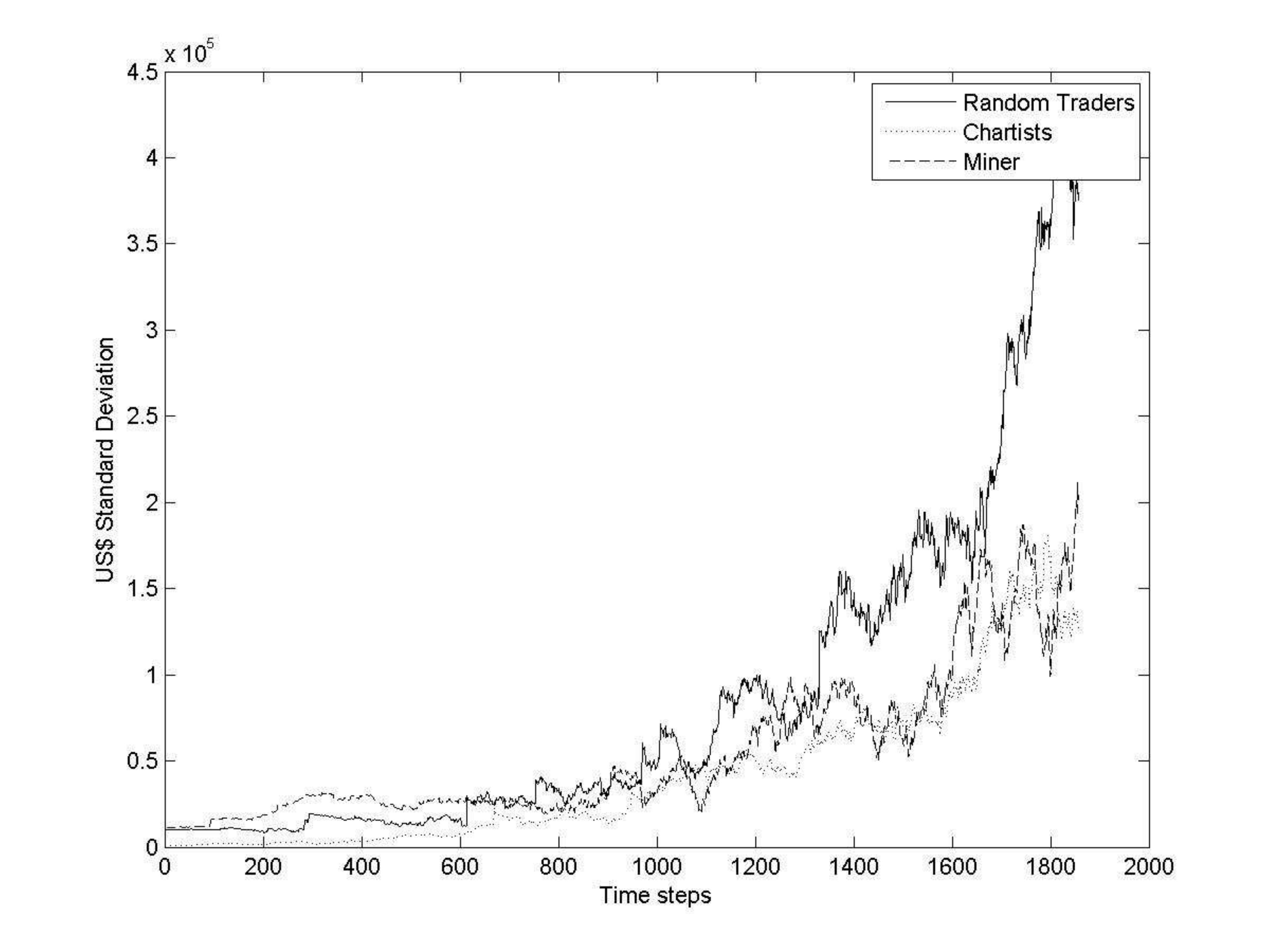}}
\caption{(a) Average and (b) standard deviation of the cash amount for all trader populations during the simulation period across all Monte Carlo simulations. \label{fig:AvgStdCash}}
\end{figure}

\begin{figure}[!ht]
\centering
\subfigure[]{
\includegraphics[width=0.45\textwidth]{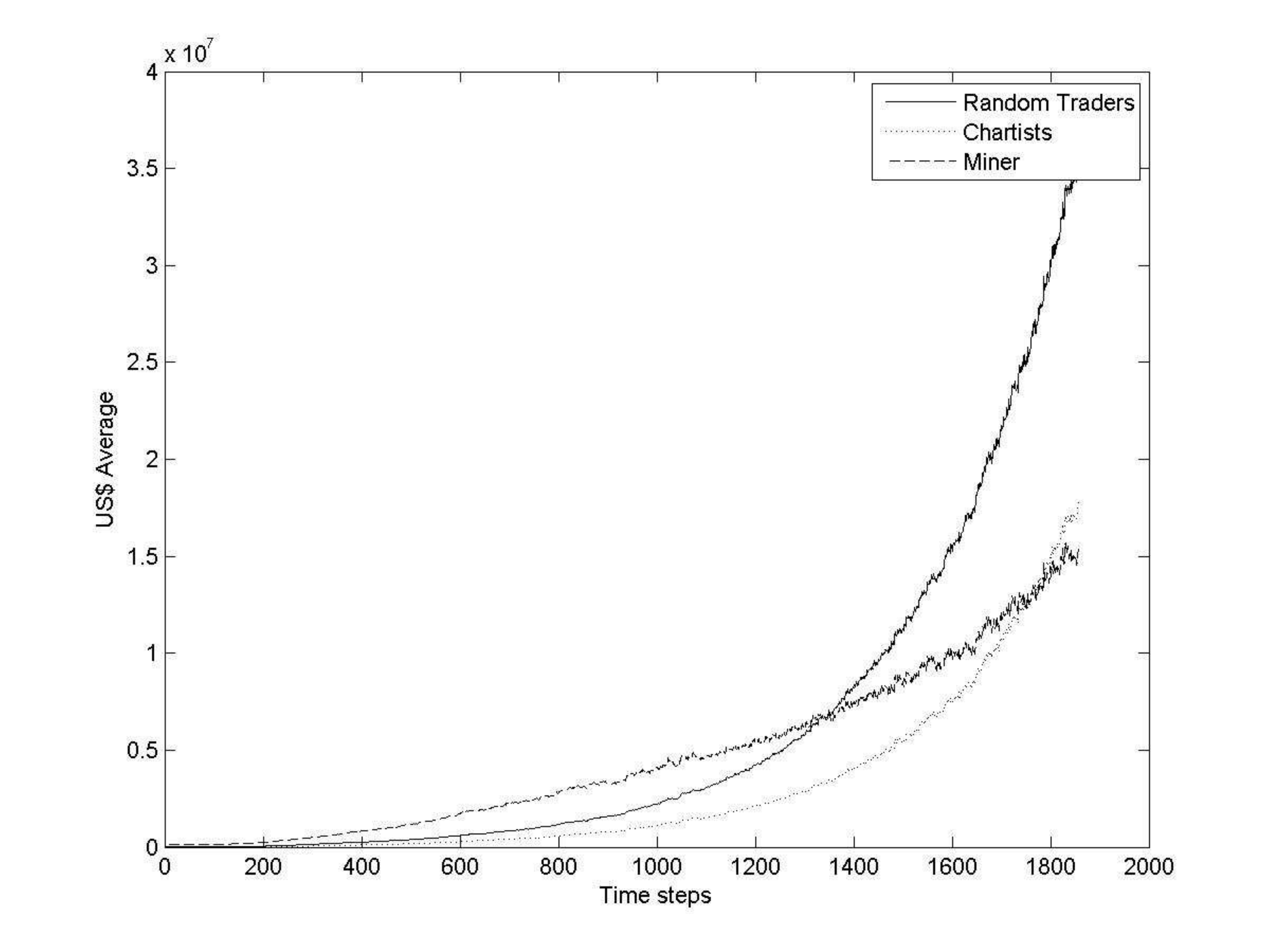}}
\hspace{4mm}
\subfigure[]{
\includegraphics[width=0.45\textwidth]{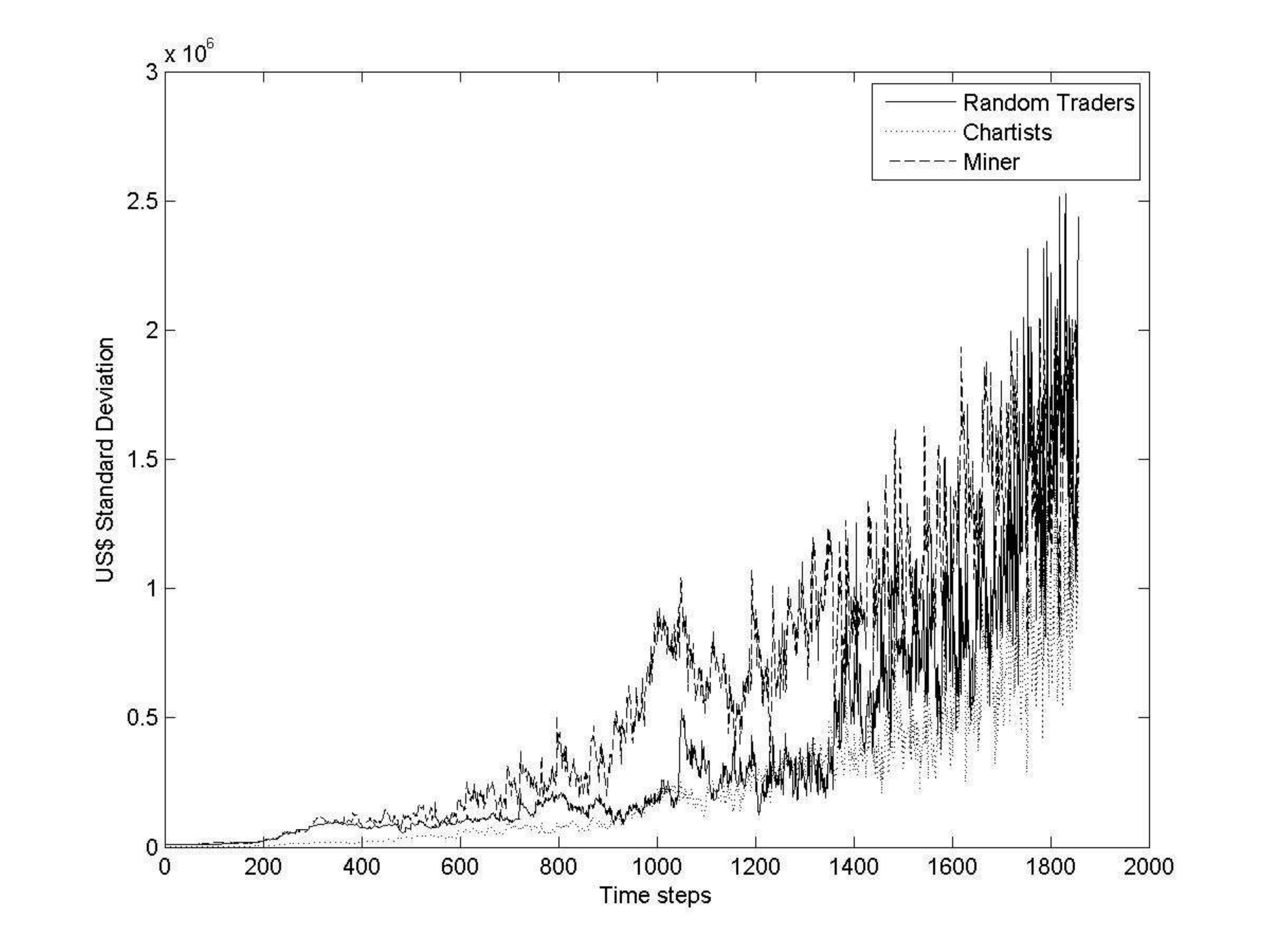}}
\caption{(a) Average and (b) standard deviation of the total wealth for all trader populations during the simulation period across all Monte Carlo simulations. \label{fig:AvgStdTotCash}}
\end{figure}

Figs. \ref{fig:AvgStdBtc} - \ref{fig:AvgStdTotCash} show the average and the standard deviation of the crypto and fiat cash, and of the total wealth, $A(t)$, of trader populations, across all 100 simulations. 
These simulations were carried with miners buying new hardware using an average percentage of 15\% of their wealth, that demonstrated to be optimal.

Figure \ref{fig:AvgStdTotCash}(a) highlights how Miners represent the richest population of traders in the market in the beginning of the simulation. However, from about 1400th step onwards, Random traders become the richest population in the market. This is mainly due to the higher number of Random traders with respect to Miners. Note also that the standard deviation of the total wealth is much more variable than the former two figures. This is due to the fact that the wealth is obtained by multiplying the number of Bitcoins by their price, which is very volatile among the various simulations, as shown in Fig. \ref{fig:averagePrice}(b).

\begin{figure}[!ht]
\centering
\includegraphics[width=0.45\textwidth]{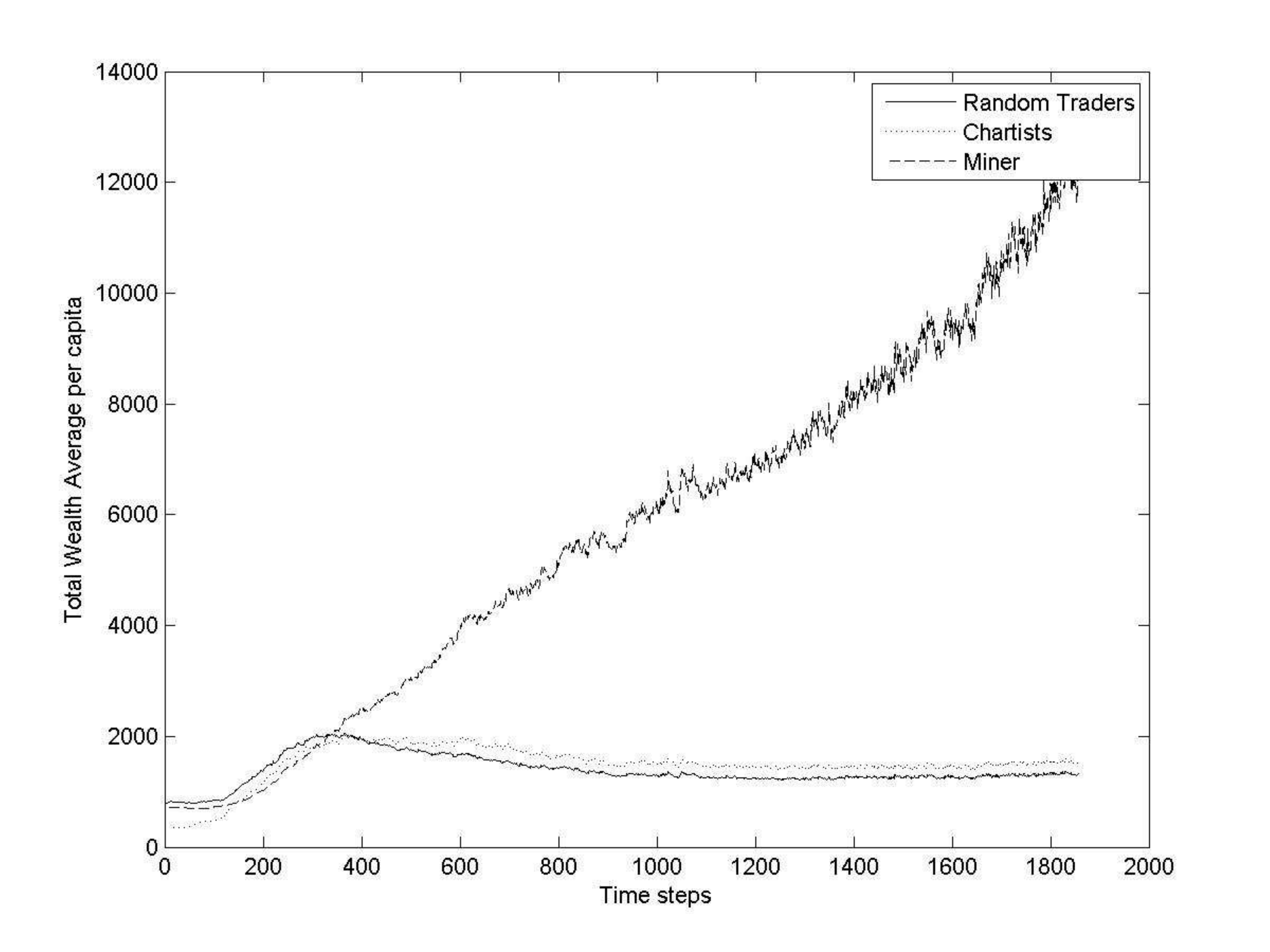}
\caption{Average across all Monte Carlo simulations of the total wealth per capita, for all trader populations. \label{fig:AvgTotCashPerCapitaOUT7}}
\end{figure}

Fig. \ref{fig:AvgTotCashPerCapitaOUT7}, shows the average of the total wealth per capita for all trader populations, across all 100 Monte Carlo simulations. Miners are clearly the winners about from the 380th simulation step onwards, thanks to their ability to mine new Bitcoins. 
Specifically, thanks to the percentage of cash that Miners devot to buy new mining hardware, Miners are able to acquire a wealth per-capite that ranges about between \$1,000 at the beginning of the simulation and \$14,000 at the end. This is due to the optimal percentage of cash devoted to buy new hardware, that is drawn from a  lognormal distribution $\gamma$ with both average and standard deviation set to  0.15, as already mentioned in \textit{The Agents}.

We varied the average percentage of their wealth that Miners devote for buying new hardware, $\gamma$, to verify how this variation can impact on Miners' success. Remember that the actual percentage for a given Miner is drawn from a log-normal distribution, so these percentages are fairly different among Miners.

\begin{figure}[!ht]
\centering
\subfigure[]{
\includegraphics[width=0.45\textwidth]{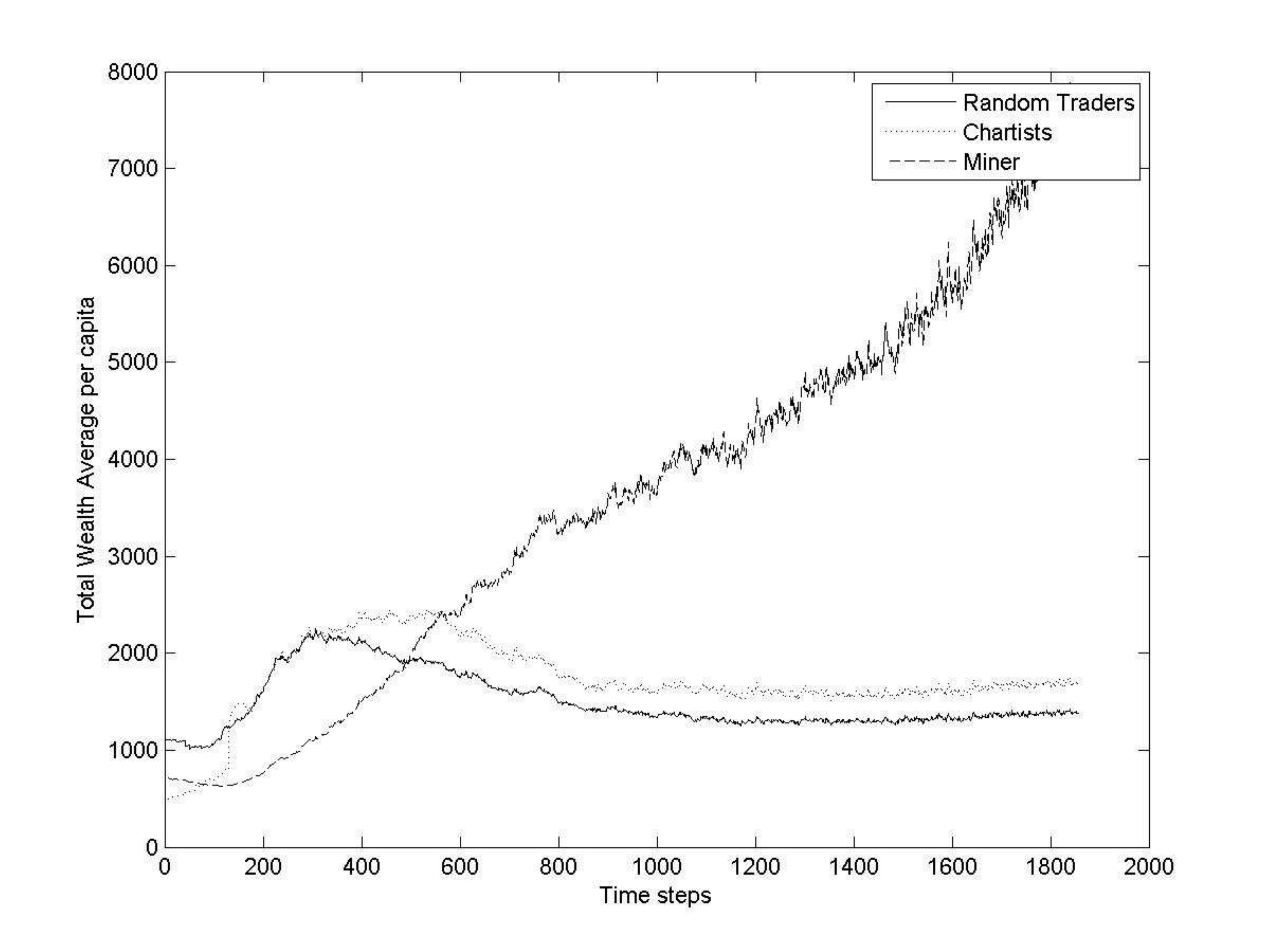}}
\hspace{4mm}
\subfigure[]{ 
\includegraphics[width=0.45\textwidth]{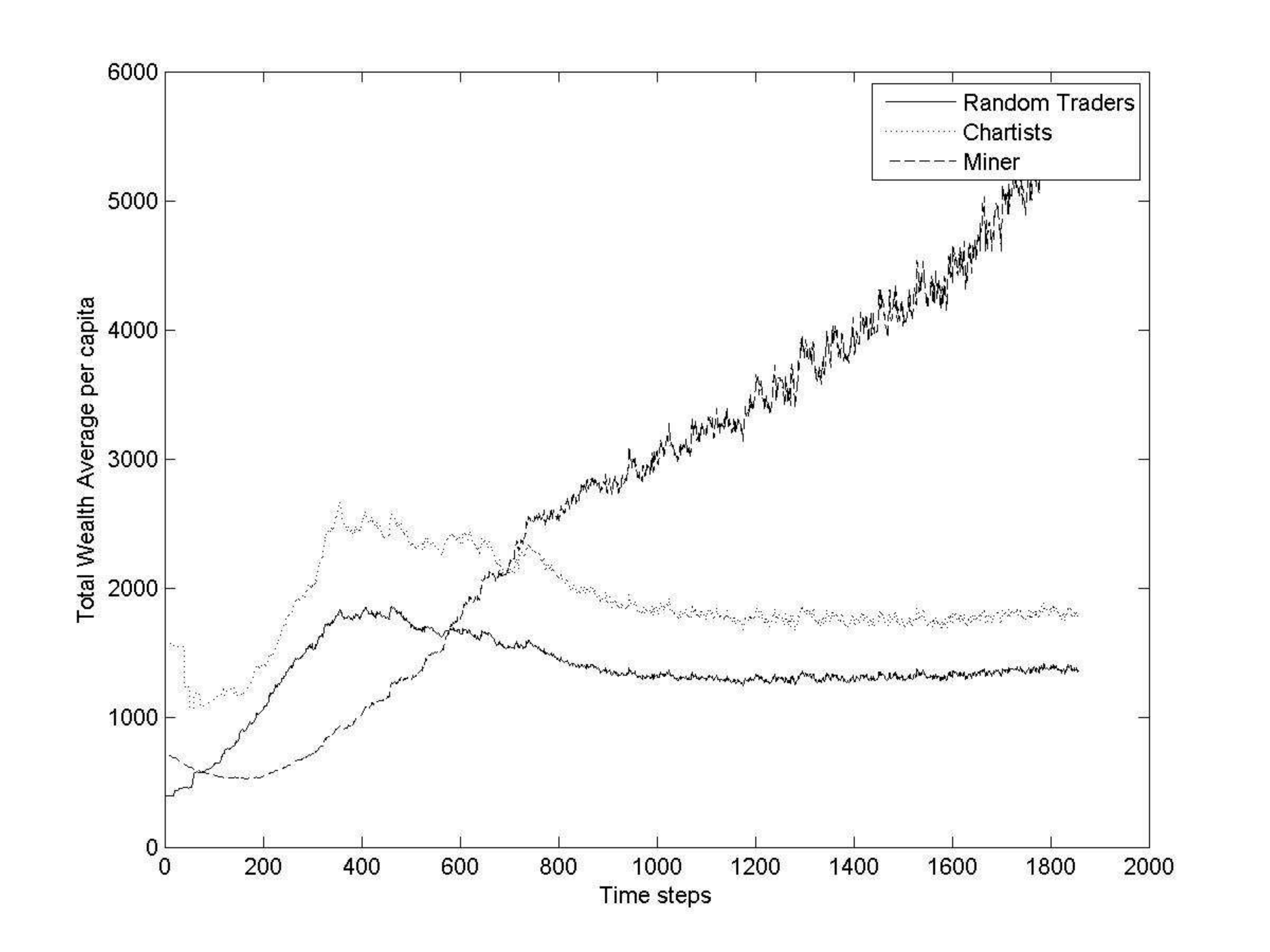}}
\caption{Average across all Monte Carlo simulations of the total wealth average per capita for all trader populations (a) for $\gamma=0.25$ and (b) for $\gamma=0.35$. \label{fig:AvgTotCashPerCapita48}}
\end{figure}

Figures \ref{fig:AvgTotCashPerCapita48} (a) and (b) show the total wealth per capita for Miners, for increasing values of the average of $\gamma$. It is apparent that Miners' gains are inversely proportional to $\gamma$, so the general strategy of devoting more money to buy hardware is not successful for Miners. 
This is because if all Miners devote an increasing amount of money to buy new mining hardware, the overall hashing power of the network increases, and each single Miner does not obtain the expected advantage of having more hash power, whereas the money spent on hardware and energy increases.
The wealth per-capite ranges between about \$1,000 at the beginning of the simulation and \$8,000 at the end, for $\gamma=0.25$ (see fig. \ref{fig:AvgTotCashPerCapita48} (a)) and about between \$1,000 and \$6,000 for $\gamma=0.35$ (see fig. \ref{fig:AvgTotCashPerCapita48} (b)).

\begin{figure}[!ht]
\centering
\subfigure[]{
\includegraphics[width=0.45\textwidth]{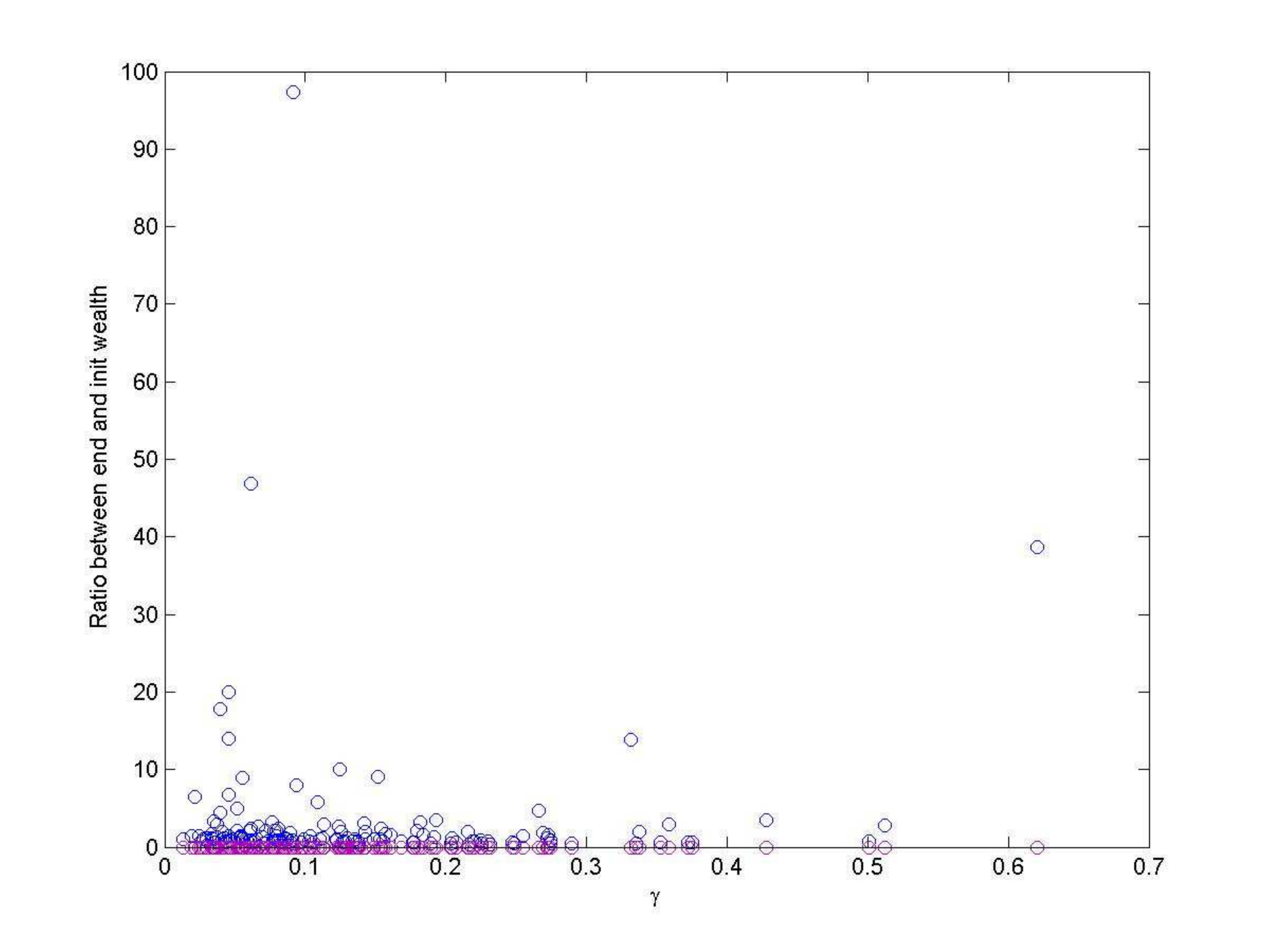}}
\hspace{7mm}
\subfigure[]{
\includegraphics[width=0.45\textwidth]{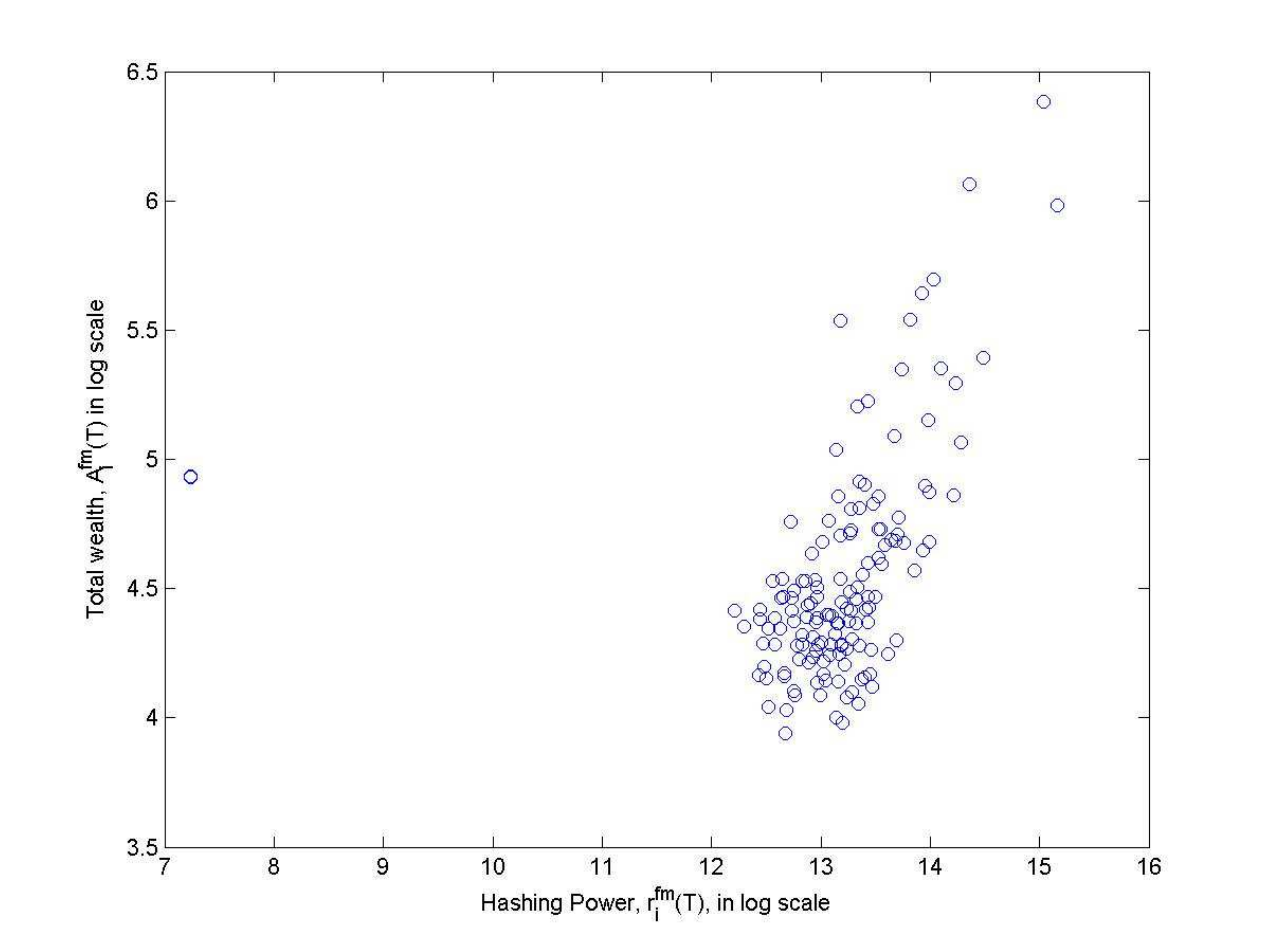}}
\caption{Scatterplots of (a) the increase in wealth of single Miners versus their average wealth percentage used to buy mining hardware, and (b) the total wealth of Miners versus their hashing power at the end of the simulation. \label{fig:crossCorr}}
\end{figure}

Having found that Miners' wealth decreases when too much of it is used to buy new hardware, we studied if increasing money spent in mining hardware is a successful strategy for single Miners, when most other Miners do not follow it.
Fig. \ref{fig:crossCorr} (a) shows the ratio of initial Miners' total wealth computed at the end and at the beginning of a single simulation, $\frac{A_i^{f_m}(T)}{A_i^{f_m}(0)}$, versus their actual value of $\gamma_i$, that is their propension to spend money to buy mining hardware. The average $<\gamma> = 0.15$ in this simulation. 
The correlation coefficients is equal to -0.14, so it looks that there is no meaningful correlation between mining success and the propension to invest in hardware. In Fig. \ref{fig:crossCorr} we can see that two of the three most successful Miners, able to increase their wealth of about 100 and 45 times, have a very low value of 
$\gamma_i$, (less than 0.1), whereas the third one, who was able to increase his wealth forty times, has a high propension to invest ($\gamma_i \simeq 0.62$).

On the contrary, we found that the total wealth, $A_i^{f_m}(T)$, of the miners at the end of the simulation is correlated with their hashing capability $r_i^{f_m}(T)$, being the correlation coefficient equal to 0.788, as shown in Fig. \ref{fig:crossCorr} (b). This result is not unexpected because wealthy Miner can buy more hardware, that in turn helps them to increase their mined Bitcoins.

\begin{figure}[!ht]
\centering
\subfigure[]{
\includegraphics[width=0.45\textwidth]{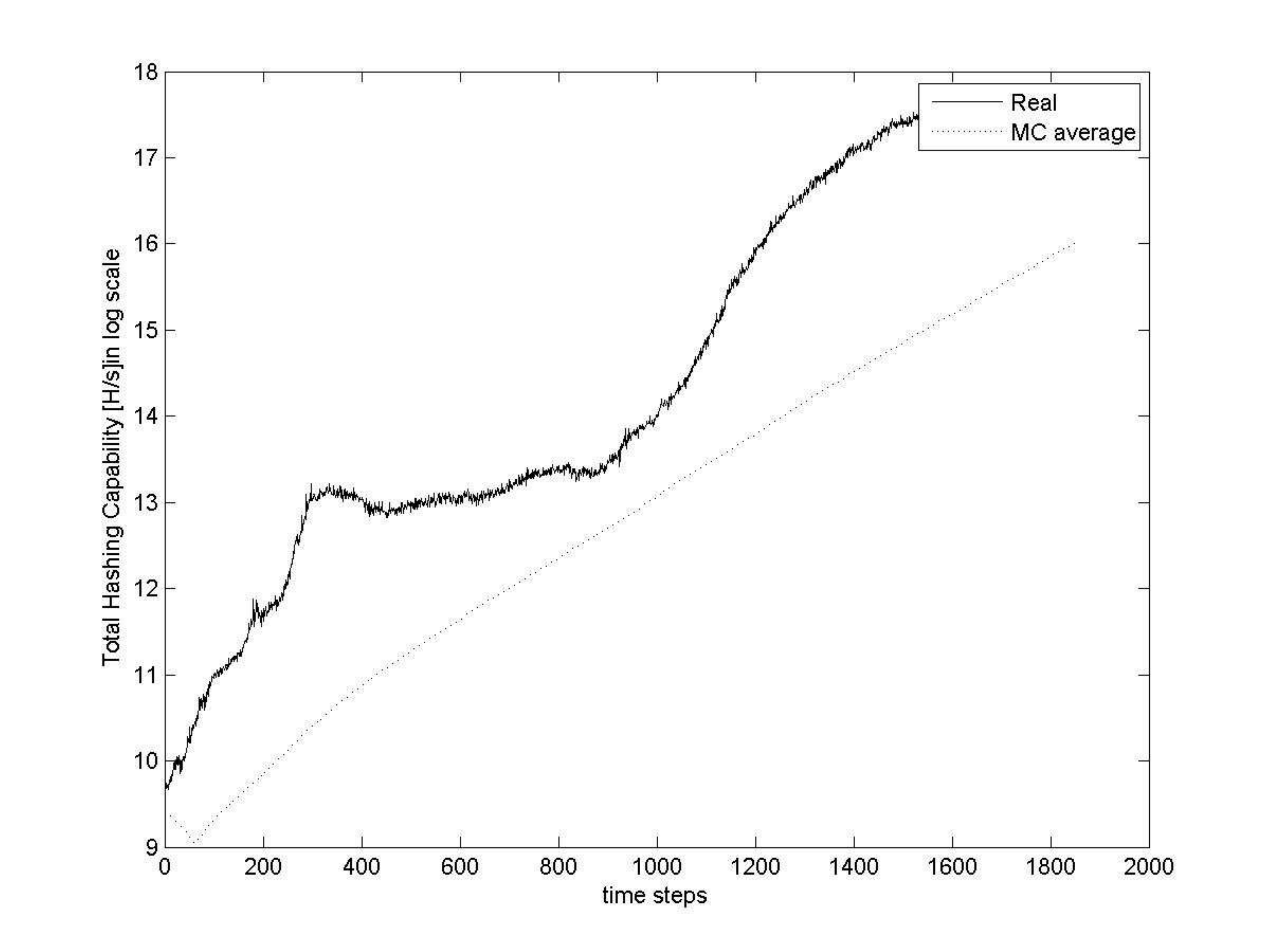}}
\hspace{7mm}
\subfigure[]{
\includegraphics[width=0.45\textwidth]{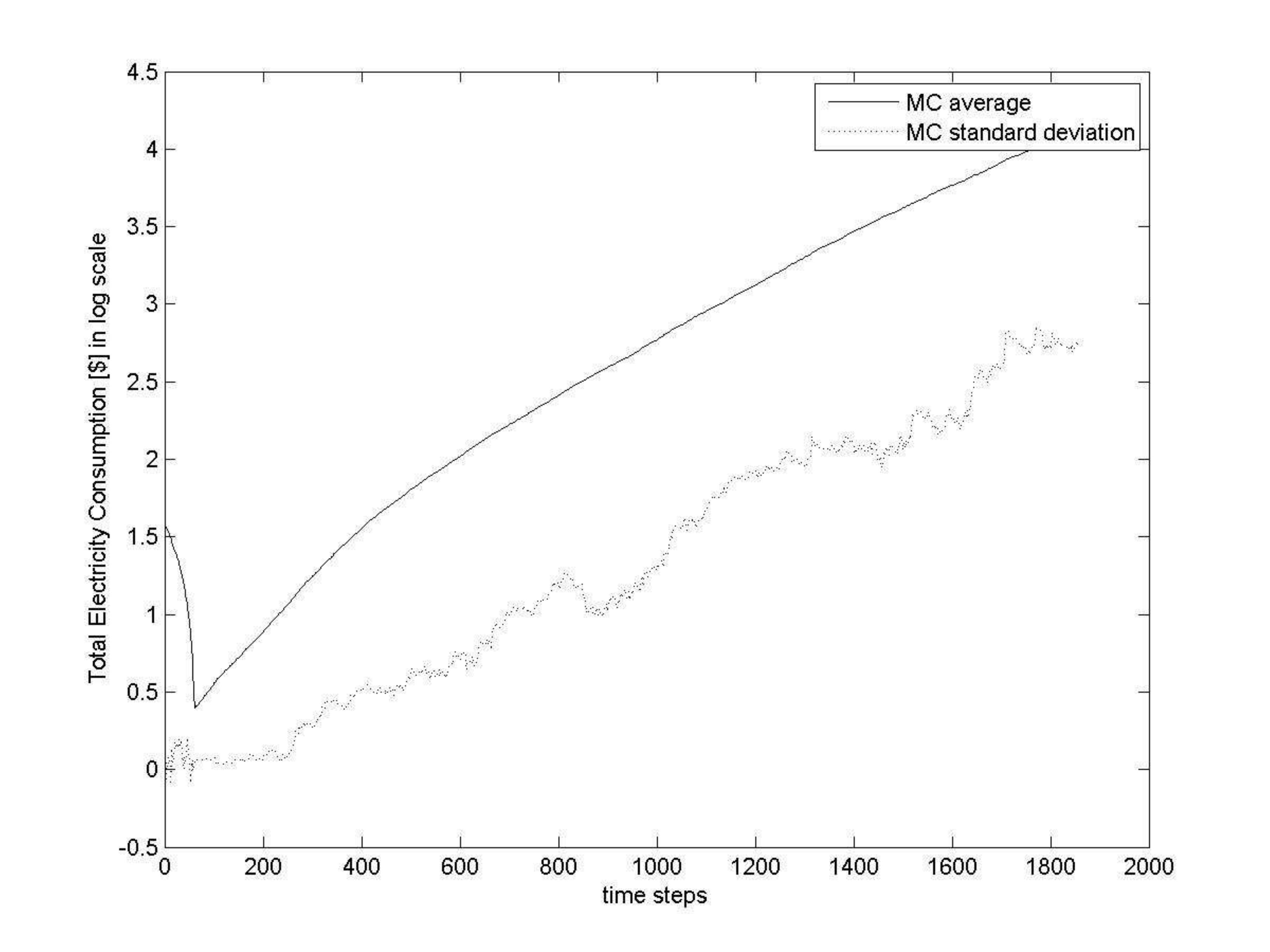}}
\caption{(a) Comparison between real hashing capability and average of the simulated hashing capability across all Monte Carlo simulations in log scale, and (b) average and standard deviation of the total expenses in electricity across all Monte Carlo simulations in log scale. \label{fig:HR}}
\end{figure}

\begin{figure}[!ht]
\centering
\subfigure[]{
\includegraphics[width=0.45\textwidth]{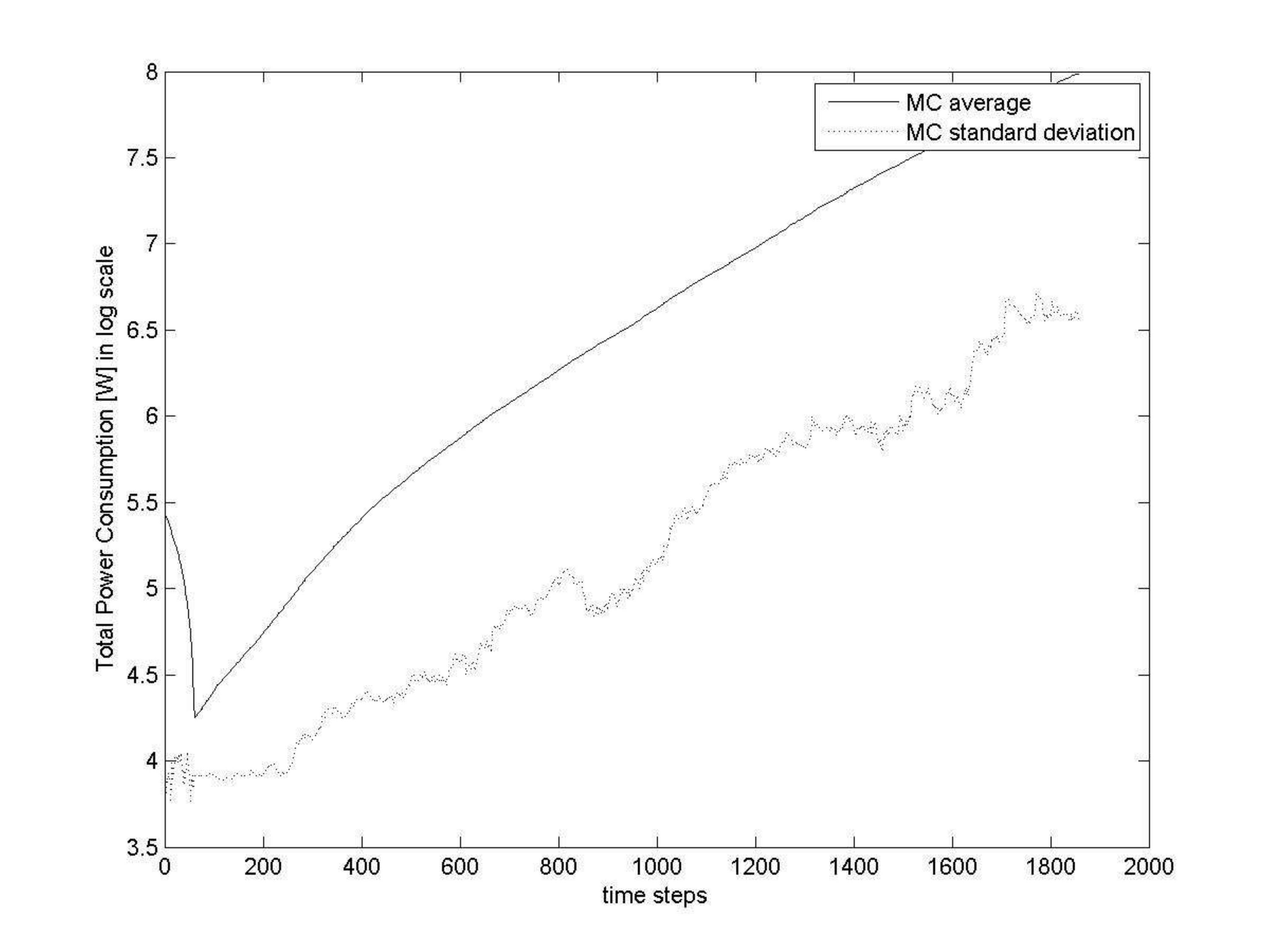}}
\hspace{7mm}
\subfigure[]{
\includegraphics[width=0.45\textwidth]{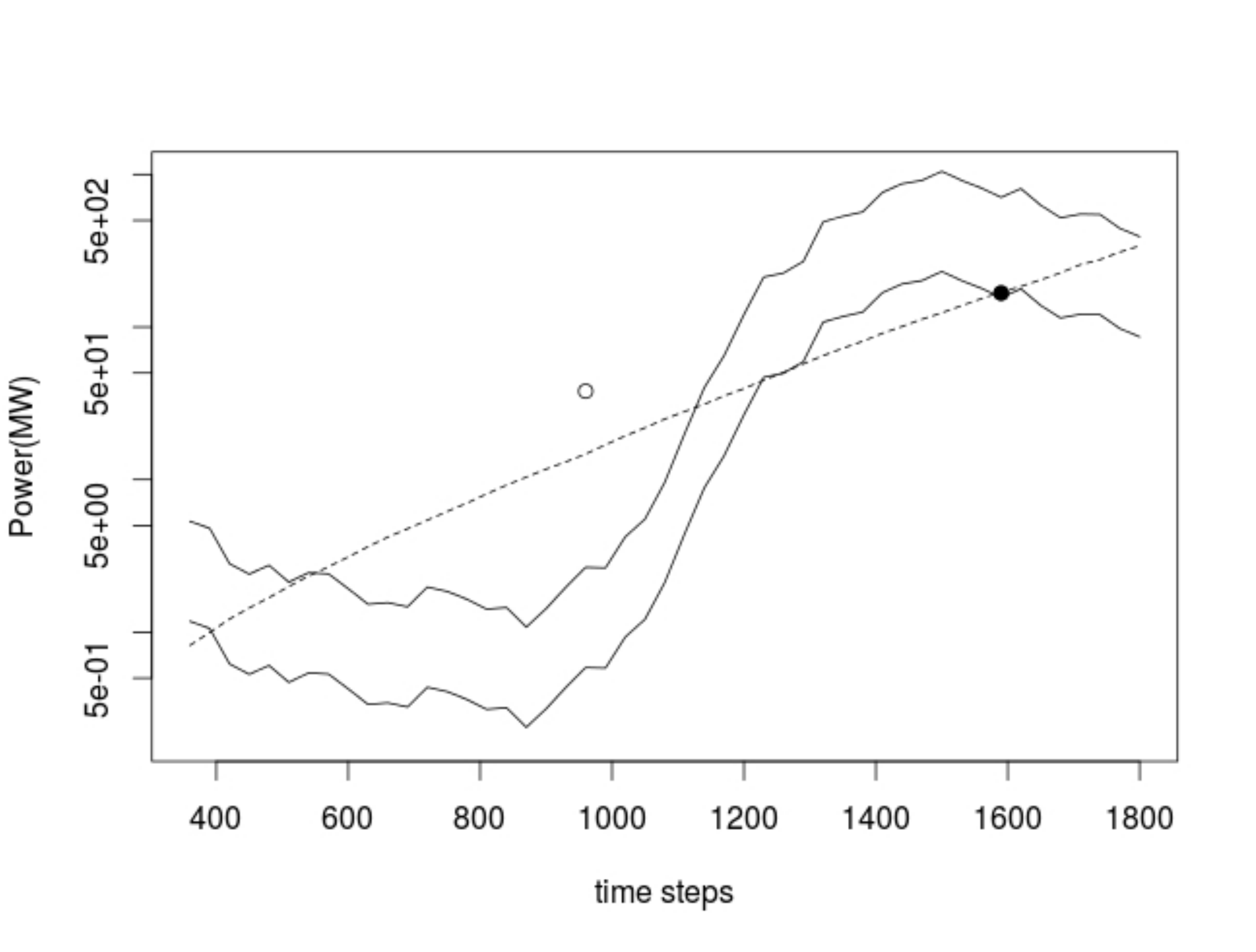}}
\caption{ (a) Average and standard deviation of the power consumption across all Monte Carlo simulations, and (b) Estimated minimum and maximum power consumption of the real Bitcoin Mining Network (solid lines), and
average of the power consumption across all Monte Carlo simulations, multiplied by 100,
the scaling factor of our simulations (dashed line). For the meaning of the circles, see text. \label{fig:electExpenses}}
\end{figure}

\begin{figure}[!ht]
\centering
\subfigure[]{
\includegraphics[width=0.45\textwidth]{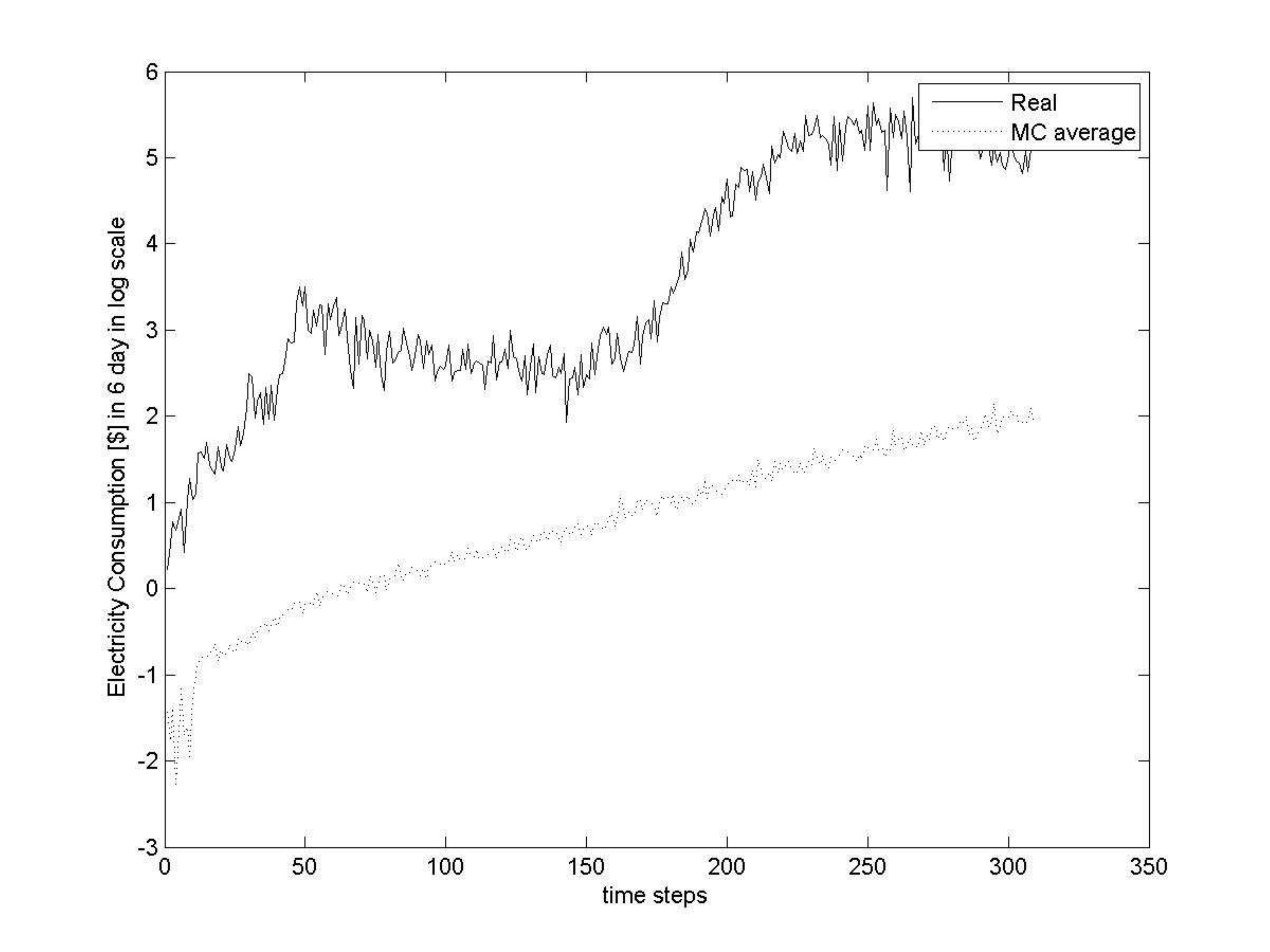}}
\hspace{7mm}
\subfigure[]{
\includegraphics[width=0.45\textwidth]{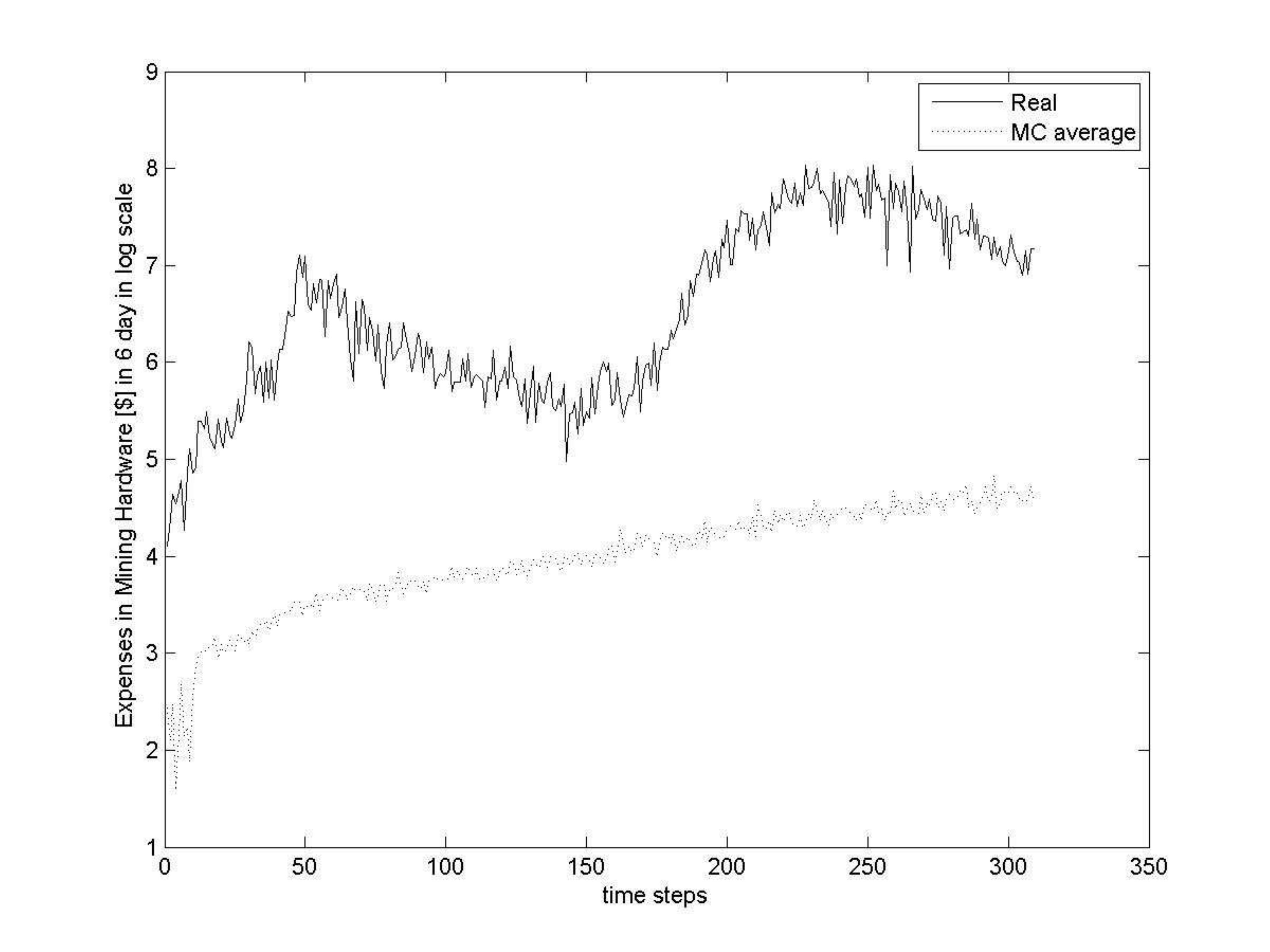}}
\caption{Real expenses and expenses average in electricity across all Monte Carlo simulations (a) and real expenses and expenses average in hardware across all Monte Carlo simulations every six days (b). \label{fig:AVGelectHardwExpenses}}
\end{figure}

\begin{figure}[!ht]
\centering
\subfigure[]{
\includegraphics[width=0.45\textwidth]{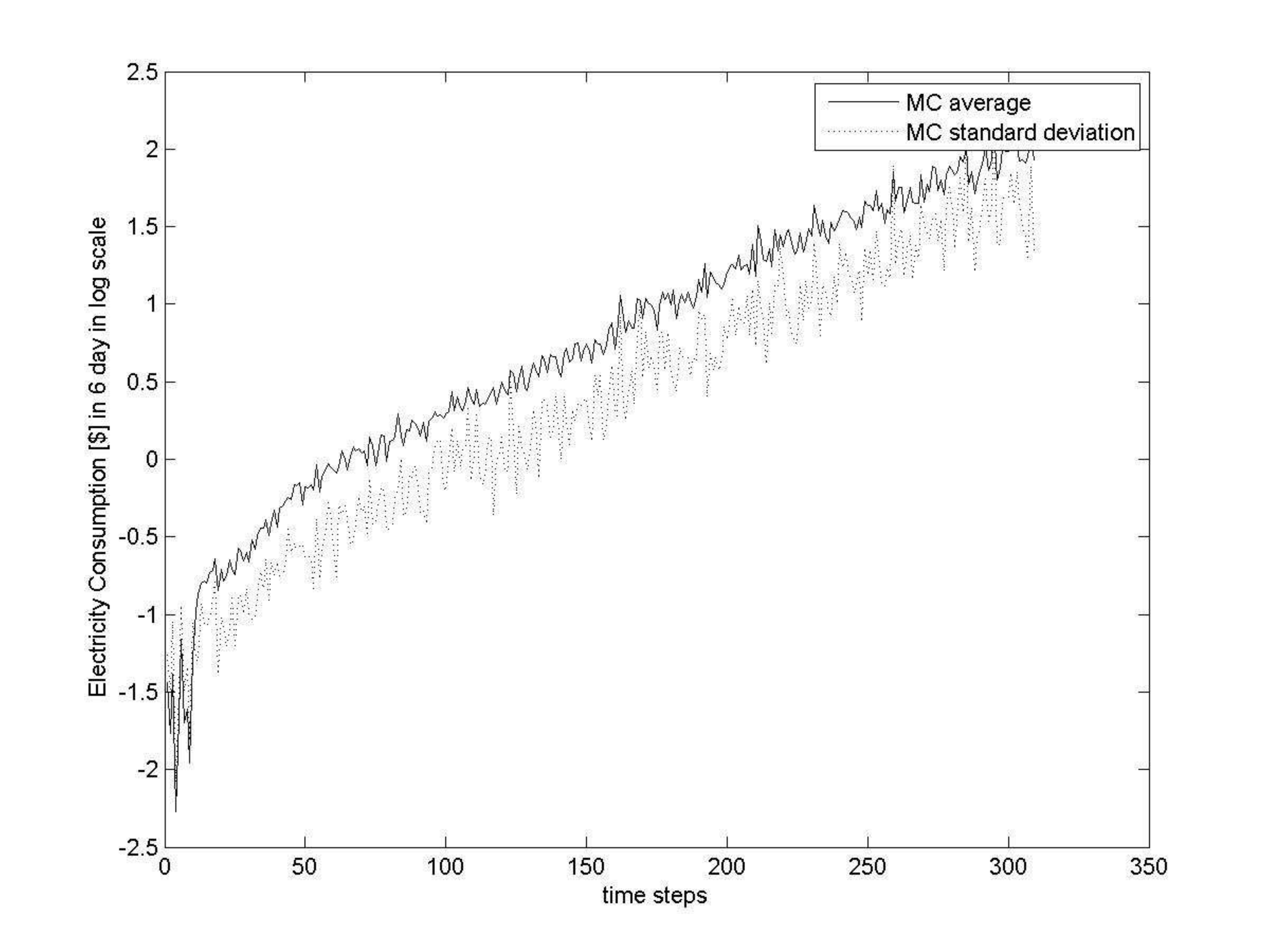}}
\hspace{7mm}
\subfigure[]{
\includegraphics[width=0.45\textwidth]{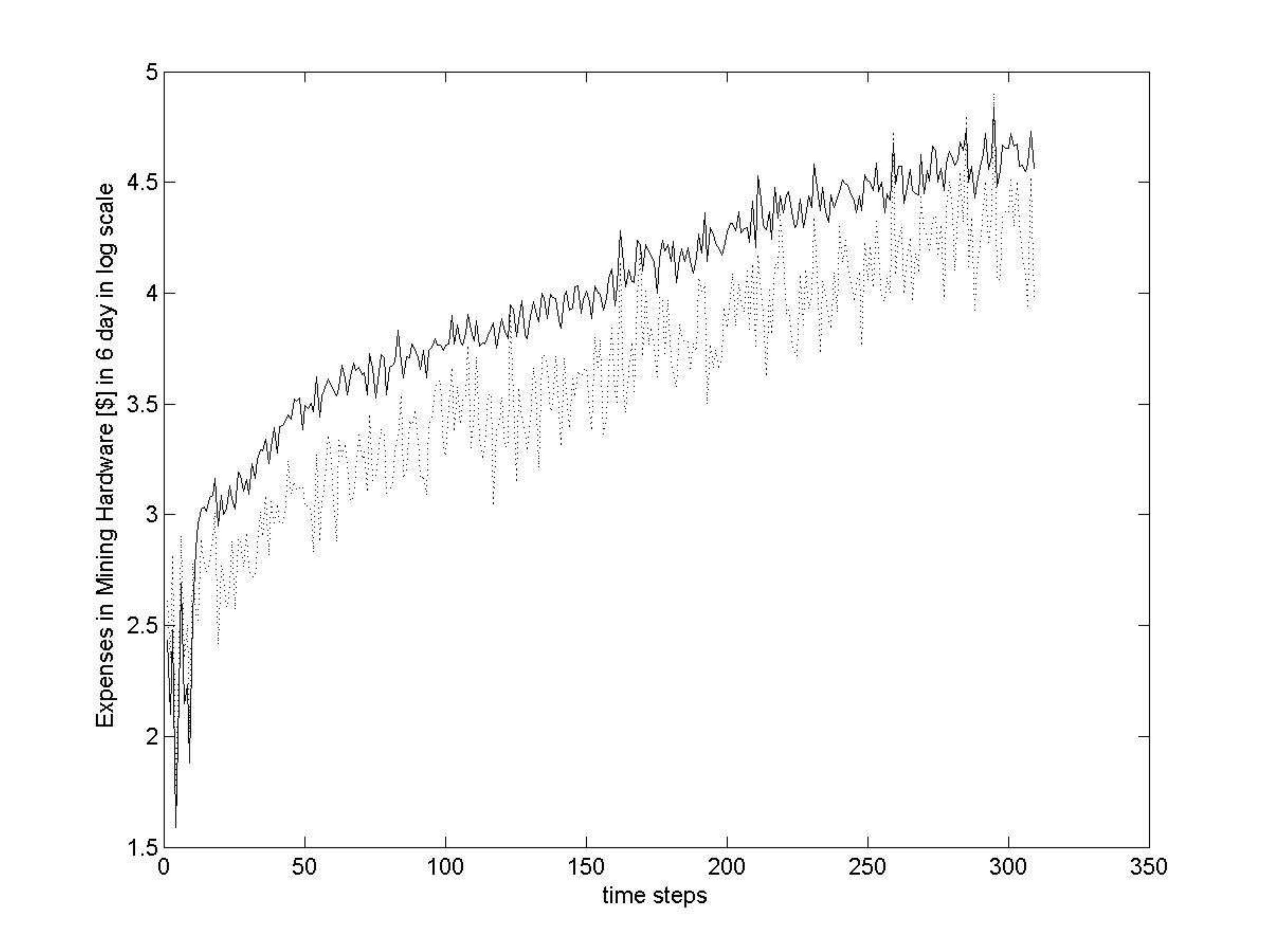}}
\caption{ Average and standard deviation of the expenses in electricity  (a)  and of the expenses in new hardware across all Monte simulations. \label{fig:STDelectHardwExpenses}}
\end{figure}

Figures \ref{fig:HR} - \ref{fig:STDelectHardwExpenses} show some significant quantities related to the Miner's population.

Fig. \ref{fig:HR}(a) shows the average hashing capability of the whole network in the simulated market across all Monte Carlo simulations and  the hashing capability in the real market, being both these quantities expressed in log scale. Note that the simulated hashing capability should be about 100 times lower than the real one, due to the reduced dimension of the simulated market with respect to the real one.
The simulated hash rate does not follow the upward trend of the Bitcoin price at about 1200th time step that is due to an exogenous cause (the step price increase at the end of 2013), that is obviously not present in our simulations. However, in Fig. \ref{fig:HR}(a) the simulated hashing capability is actually about two orders of magnitude lower than the real one, as it should be.

In general, Bitcoin mining hardware become obsolete from a few months to one year after you purchase them. "Serious" miners usually buy new equipment every month, re-investing their profits into new mining equipment, if they want that their Bitcoin mining operation to run long term (see web site http://coinbrief.net/profitable-bitcoin-mining-farm/.
In our model, miners divest their mining equipment about every ten months.

Figure \ref{fig:electExpenses} (a) shows the average and standard deviation of the power consumption across all Monte Carlo  simulations. 
Figure \ref{fig:electExpenses} (b) shows an estimated minimum and maximum power consumption of the Bitcoin mining network, together with the average of the power consumption of Fig. \ref{fig:electExpenses} (a), in logarithmic scale. 
The estimated theoretical minimum power consumption is obtained by multiplying the actual hash rate of the network at time $t$ (as shown in Fig. \ref{fig:HR}(a)) with the power consumption $P(t)$ given in eq. \ref{P}. This would mean that the entire hashing capability of miners is obtained with the most recent hardware.
The estimated theoretical maximum power consumption is obtained by multiplying the actual hash rate of the network with the power consumption $P(t-360)$, referring to one year before. This would mean that the entire hashing capability of miners is obtained with hardware one year old, and thus less efficient. The estimated obsolescence of mining hardware is between six months and one year, so the period of one year should give a reliable maximum value for power consumption.

The simulation results, averaged on 100 simulations, show a much more regular trend, steadily increasing with time -- which is natural due to the absence of external perturbations on the model. However, the power consumption value is of the same order of magnitude of the "real" case. Note that the simulated consumption shown in Fig. \ref{fig:electExpenses} (b) is multiplied by 100, that is the scaling factor of our simulations, that have $1/100^{th}$ of the real number of Bitcoin traders and miners.

Fig. \ref{fig:electExpenses} (b) also shows a white circle, at time step corresponding to April 2013, with a value of 38.8 MW. This value has been taken by Courtois et al, who in work \cite{CourtoisGrajek} write:
\begin{quote}
\small{In April 2013 it was estimated that Bitcoin miners already used about 982 Megawatt hours every day. At that time the hash rate was about 60 Tera Hash/s. (Refer to article by Gimein Mark "Virtual Bitcoin Mining Is a Real-World Environmental Disaster", 13 April 2013 published on web site www.Bloomberg.com.).
}
\end{quote}

In fact, the hash rate quoted is correct, but the consumption value looks overestimated of one order of magnitude, even with respect to our maximum power consumption limit. We believe this is due to the fact that the authors still referred to FPGA consumption rates, not fully appreciating how quickly the ASIC adoption had spread among the miners.

As of 2015, the combined electricity consumption was estimated equal to 1.46 Tera Wh per year, that corresponds to about 167 MW (see article "The magic of mining", 13 January 2015 published on web site www.economist.com.). 
This value is reported in Fig. \ref{fig:electExpenses} (b) as a black circle. 
This time, the value is slightly underestimated, being at the lower bound of power consumption estimate, and 
is practically coincident with the average value of our simulations.

Figures \ref{fig:AVGelectHardwExpenses} (a) and (b) show an estimate of the expenses incurred every six days in electricity (a) and in hardware (b) for the new hardware bought each day in the real and simulated market. 

Note that also the values of the simulated expenses are average values across all Monte Carlo simulations.

These expenses were computed assuming that the new hardware bought each day in the real (simulated) market, and hence the additional hashing capability acquired each day, is equal to the difference between the real (simulated) hash rate in $t$ and the real (simulated) hash rate in $t-1$.

For both these expenses, contrary to what happens to the respective real quantities, the simulated quantities do not follow the upward trend of the price due to the constant investment rate in mining hardware.

Figure \ref{fig:STDelectHardwExpenses} (a) and (b) show the average and standard deviation, across all Monte simulations, of the expenses incurred every six days in electricity and in new hardware respectively, showing the level of the variation across the simulations.

Remembering that our model sizes the artificial market at about 1/100 of the real market and that the number of traders, their cash and their trading probabilities are rough estimates of the real ones, the simulated market outputs can be considered reasonably close to the real one. 

\subsection{Other Results}\label{sec:6.3}

It is known that Bitcoin price is driven by speculation, government regulation and investors behavior, and its volatility depends also on Bitcoin acceptance and usage.
In 2012 and 2013 prices had a wild ride, until they reached a peak of \$1,150 in December 2013. In 2014 Bitcoin price fell following the shutdown of historical Mt. Gox exchange site and reports regarding Bitcoin ban in China. 

Trying to reproduce this market trend, we introduced in the model a particular speculative behaviour by some traders.
The speculative mechanism implemented stems from a report, called the "Willy Report", published by an anonymous researcher, which alleges suspicious trading activity at Mt. Gox. "The Willy Report: proof of massive fraudulent trading activity at Mt. Gox, and how it has affected the price of Bitcoin", was posted on May 25, 2014 in web site https://willyreport.wordpress.com/. 

The anonymous researcher claims to have noted a suspicious bot behavior on Mt. Gox, that spread its trading activity over many accounts, and how this fraudulent massive trading activity impacted on the price, causing bubble and crash.

\vspace{0.5cm}
In the report the researcher writes: 

\begin{quote}
\small{"Somewhere in December 2013, a number of traders including myself began noticing suspicious bot behavior on Mt. Gox. Basically, a random number between 10 and 20 bitcoin would be bought every 510
minutes, nonstop, for at least a month on end until the end of January. The bot was dubbed “Willy” \dots its trading activity was spread over many accounts. \dots  Their trading activity went back all
the way to September 27th.\dots 

In total, a staggering about \$112 million was spent to buy close to 270,000 BTC - the bulk of which was bought in
November. So if you were wondering how Bitcoin suddenly appreciated in value by a factor of 10 within the span of one month, well, this may be why. Not Chinese investors, not the Silkroad bust - these events may have
contributed, but they may not have been main reason. \dots 

\dots there was another timetraveller account with an ID of 698630 - and this account, after being active for close to 8 months, became completely inactive just 7 hours before the first Willy account became active! So it is a reasonable assumption that these accounts were controlled by the same entity. \dots There were several peculiar things about Markus. First, its fees paid were always 0 (unlike Willy, who paid fees as usual). Second, its fiat spent when buying coins was all over the place, with seemingly completely random prices paid per bitcoin.\dots 

\dots Since there are no logs past November 2013, the following arguments are largely based on personal speculation, and that of other traders\dots

on January 26th, Willy suddenly became inactive -- and with it, the price retraced back to a more reasonable spread with the other exchanges. Shortly after -- on February 3rd to be precise -- it seemed as if Willy had begun to run in reverse, although with a slightly altered pattern: it seemed to sell around 100 BTC every two hours. \dots There's some additional evidence on the chart that a dump bot may have been at play. At several points in time, starting from Feb. 18th, it seemed that some bot was programmed to sell down to various fixed price levels.\dots 

At this point, I guess the straightforward conclusion would be that this is how the coins were stolen: a hacker gained access to the system or database, was able to assign himself accounts with any amount of USD at will, and just started buying and withdrawing away.\dots"
}.
\end{quote}

According to what just mentioned, we modeled a similar behaviour. We assumed that, until the end of January 2014, 40\% of Random traders entering the market were Mt. Gox accounts. 
The  Mt. Gox accounts have a behaviour equal to that of Random traders described in Paragraph \textit{Random Traders} until July 2012. Then, from August 2012 and until the end of January 2014, they issue only buy orders. Next, from February 2014 they issue only sell orders. Their trading probability is set equal to 0.1 in every period.

Fig. \ref{fig:Price} shows the Bitcoin price in one typical simulation run, under these conditions. 
At first, the price keeps its value constant, then at about 700 simulation steps, it grows as happens in reality.
The price maintains its value high for about 500 simulation steps, then its value falls down, but after a short delay it continues on its upward slope until the end of the simulation, due to the intrinsec mechanism of our model already previous described.

\begin{figure}
\centering
 \includegraphics[width=0.5\textwidth]{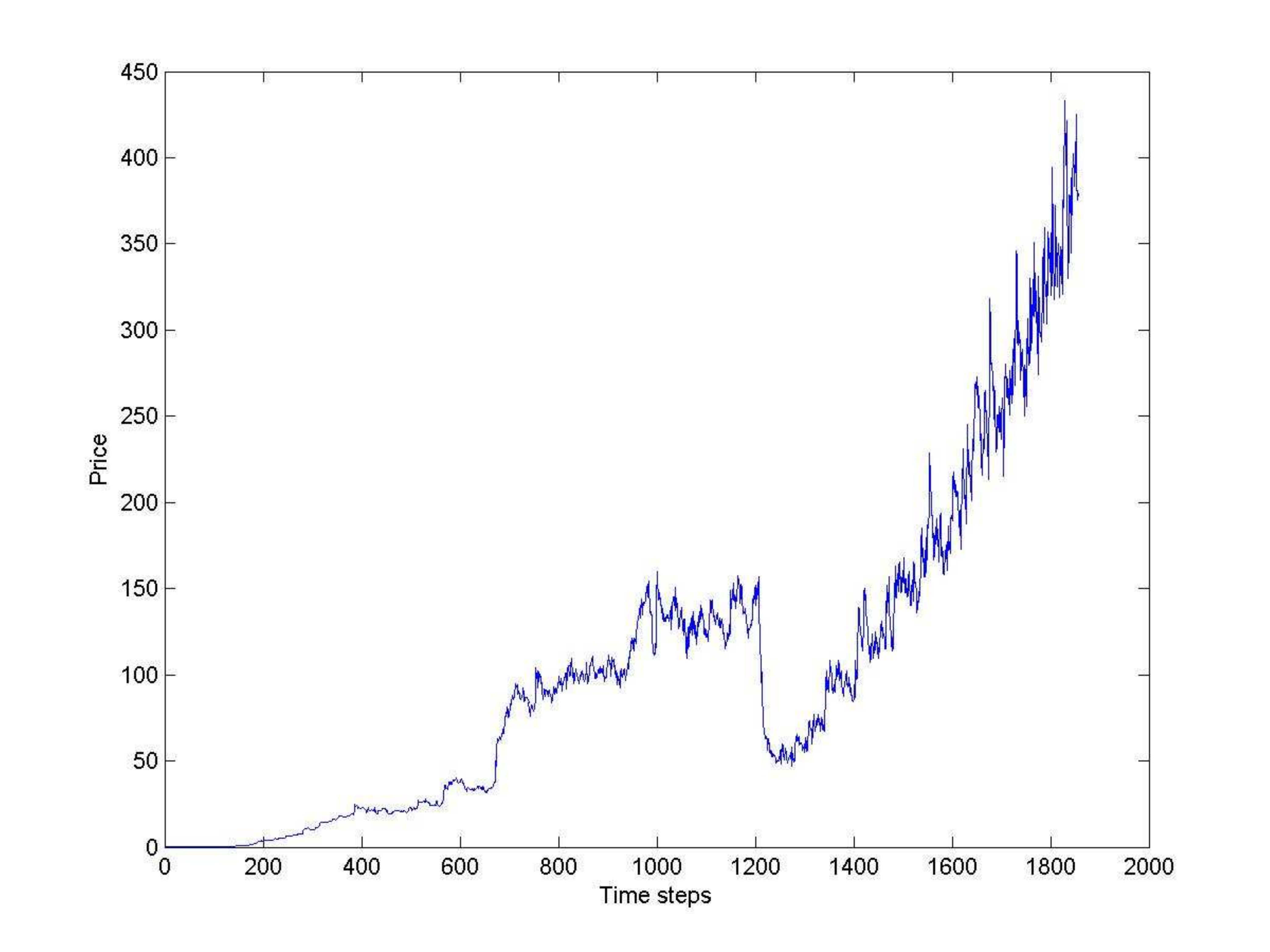}
\caption{Price of the Bitcoin in the simulated market.}
\label{fig:Price}       
\end{figure} 

The MtGox accounts' behaviour has a key rule in the reproduction of the price that has a trend more similar to the real one (shown in Fig. \ref{fig:realPrice}) than that described in section \textit{Bitcoin prices in the real and simulated market}.

Figs. \ref{fig:averagePrice} (a) and (b) report the average and the standard deviation of the simulated price across all Monte Carlo simulations showing the consistency of the results.

\begin{figure}[!ht]
\centering
\subfigure[]{
\includegraphics[width=0.45\textwidth]{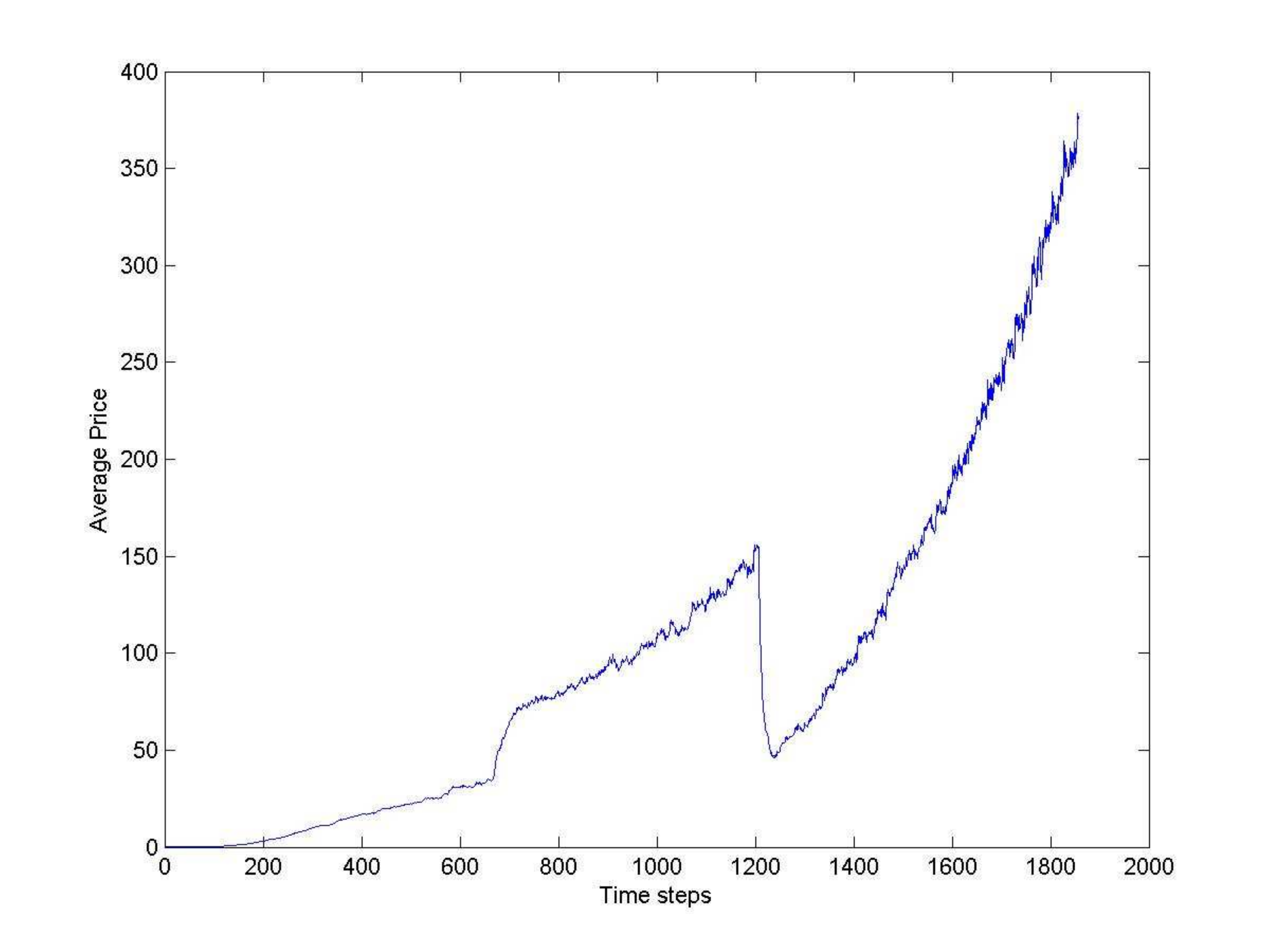}}
\hspace{7mm}
\subfigure[]{
\includegraphics[width=0.45\textwidth]{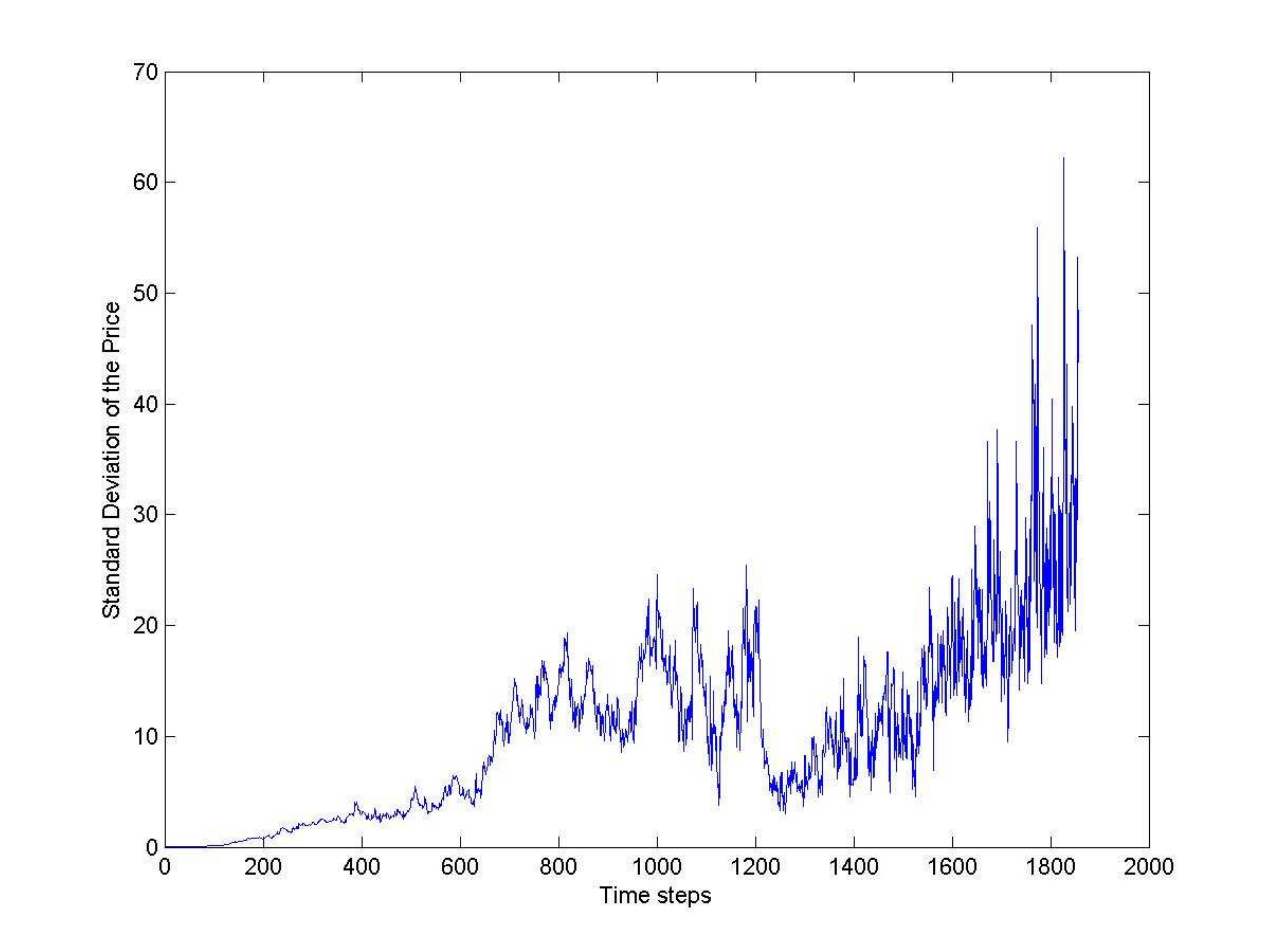}}
\caption{(a) Average Price  and (b) standard deviation computed on the 100 Monte Carlo simulations performed. \label{fig:averagePrice}}
\end{figure}

All the analysis described in the previous sections was performed also for the model including "Mt.Gox" accounts, producing results consistent with those studied in the previous sections.

\section{Conclusions} \label{sec:7}
In this work, we propose an heterogenous agent model of the Bitcoin market with the aim to study and analyse the mining process and the Bitcoin market starting from September 1st, 2010, the approximate date when miners started to buy mining hardware to mine Bitcoins, for five years. 

The proposed model simulates the mining process and the Bitcoin transactions, by implementing a mechanism for the formation of the Bitcoin price, and specific behaviors for each typology of trader. 
It includes different trading strategies, an initial distribution of wealth following Pareto law, a realistic trading and price clearing mechanism based on an order book, the increase with time of the total number of Bitcoins due to mining, and the arrival of new traders interested in Bitcoins. 
 
The model was simulated and its main outputs were analysed and compared to respective real quantities with the aim to demonstrate that an artificial financial market model can reproduce the stylized facts of the Bitcoin financial market.

The main result of the model is the fact that some key stylized facts of Bitcoin real price series and of Bitcoin market are very well reproduced. 
Specifically, the model reproduces quite well the unit root property of the price series, the fat tail phenomenon, the volatility clustering of the price returns, the price peak in November 2013, its next fall in April 2014, the generation of Bitcoins, the hashing capability, the power consumption, and the hardware and electricity expenses incurred by Miners.

The proposed model is fairly complex. It is intrinsically stochastic and of course it includes endogenous mechanisms affecting the market dynamics. 
The Zipf distribution of traders' wealth, that impacts to the size of the orders and the "herding" effect of Chartists, when a price trend is established, play a key role in the distribution of the price returns, and hence in the reproduction of the fat tail phenomenon.
The Chartist behavior and also the variability of the spread of limit
prices as a function of past price volatility contribute to the volatility clustering. 
The threshold of activation of Chartists based on price relative variation, and the past price volatility used to determine the spread of limit prices impact on the unit-root property of the price series.
The percentage of each trader's population, the choice of a trader that trades at a given time step, and the type of trading (buy or sell), as well as the setting of the quantity to trade, impact on the price trend.
The setting of the amount of cash to devote to buy new hardware impacts on the wealth and hashing capability of Miners, and consequently on their hardware and electricity expenses.

Future research will be devoted to study in deeper detail the mechanisms impacting on the model dynamics. 
In particular, we will perform a comprehensive analysis of the sensitivity of the model to the various parameters, and will add traders with more sophisticated trading strategies, to assess their profitability in the simulated market.
In addition, since the calibration of our model is based on very few specific real data, and on many assumptions aiming to derive the needed data from indirect real data, we plan to perform a deeper analysis of the Block Chain, and to gather financial data from the existing exchanges, in order to extract specific information needed for a better calibration of our model.

\appendix
\section{Appendix}\label{Appendix}
%
%

\section{Trader Wealth Endowment}\label{sec:App.1}

The distributions of cash and Bitcoins, for traders in the market at initial time, follow a power-law with exponent $\alpha$ set equal to 1, a value yielding the distribution known as Zipf's law \cite{Newman}. 
This is the same assumption made in other papers \cite{Raberto2001,Raberto2003}.
We assumed that 40\% of the total initial crypto cash is not involved in the trading activity, given the initial Miners hoarded a fraction of their Bitcoins (see web site http://www.coindesk.com/mit-report-bitcoin-more-likely-spent-hoarded/.).

To create the Zipf's distribution we used the ranking property of the power-law 
\cite{Takayasu}.  If the total number of traders is $N_t$ and the number of Bitcoins owned by them, $b_i$, follows a Pareto law with exponent $\alpha = 1$, it is well-known from Harmonic-series theory that the total number of Bitcoins $B_T = b_1 ln(N_t) + \gamma$, where $\gamma$ is the Euler-Mascheroni constant and $b_1$ is the number of Bitcoins owned by the richest trader. The number of Bitcoins owned by $i$-th trader is consequently: $\frac{b_1}{i}$.
We set the cash of each initial trader equal to five times the value of their crypto cash. 

A similar approach was followed to set the wealth of traders who enter the simulation at $t>0$, but in this case the traders are only endowed with cash. In this case, we had no specific data to calibrate the wealth of these traders. We stipulated that the cash, $c^s_1$, of the richest trader is about five times the cash owned by the richest initial trader, and that the exponent of the Pareto law is in this case $\alpha = 0.6$. Overall, we performed various simulations varying these parameters, with no significant impact on the results.

The set of "new" traders are generated before the simulation starts. When new traders are needed to enter the simulation, they are chosen randomly in this set.

\section{Active Traders}\label{sec:App.2}
As mentioned in subsection \textit{The agents}, only a given percentage of traders is active in the market, and hence enabled to issue orders. To compute this percentage we made some assumptions starting from the work \cite{Ron} from which we extracted Table \ref{TransactionNumber}, which shows the distribution of the transactions number per entity (\textit{Entity} means the common owner of multiple Bitcoin addresses) and per address on a period between January 3rd, 2009 and May $13th$ 2012.

\begin{table}[!hbt]

\begin{center}
\begin{small}
\caption{\textit{The distribution of the number of transactions per entity and per address. }\label{TransactionNumber}}
\scalebox{1}{
\begin{tabular} [t]{|p{4cm}|p{3cm}|p{3cm}|}
\hline
Number of & Number of & Number of \\
 Transactions (n)&entities &addresses\\
\hline  
$ 1 \le n <2$ &557,783 &495,773\\
\hline  
$ 2 \le n <4$ &1,615,899 &2,197,836\\
\hline  
$ 4 \le n <10$ & 222,433&780,433\\
\hline  
$ 10 \le n <100$ & 55,875&228,275\\
\hline  
$ 100\le n <1,000$ & 8,464&26,789\\
\hline  
$ 1,000 \le n <5,000$ &287 &1,032\\
\hline  
$ 5,000 \le n <10,000$ &35 &51\\
\hline  
$ 10,000 \le n <100,000$ & 32&24\\
\hline  
$ 100,000 \le n <500,000$ & 7&3\\
\hline  
$ n \ge 500,000 $ &1 &2\\
\hline
\end{tabular}
}
\end{small}
\label{tab:TransactionNumber}
\end{center}
\end{table}

By analyzing Table \ref{TransactionNumber}, we can  observed that 
97.37\% of all entities had fewer than 10 transactions each, 
2.27\% of all entities had a number of transactions ranging from 10 to 100,   0.34 \% of all entities had a number of transactions ranging from 100 to 1,000, and the remaining 0.02 \% had a number of transactions higher than 1,000.

According to the insights coming from this work, we neglected the entities having fewer than 10 transactions each and the entities having more than 1,000 transactions each. This was done hypothesizing that the formers, typically involving a small number of Bitcoins, refer to entities who made transactions by chance "only to use" this new coin, whereas the latters, involving a very high of transactions, are probably not linked to single traders, but are the addresses of exchange sites, or of retailers accepting Bitcoins.

We considered the remaining entities, that is the entities with a number of transactions in the range from 10 to 1,000.
They issue orders with a period ranging from about one day to about 122 days, because these data are computed on 1227 days.
As a result, the daily trading probability of an entity ranges from 0.008 to 1.

We set the values of trading probabilities for Random traders and Chartists in this range. Specifically, we assumed that, Random Traders  are active with a probability $p^t_R = 0.1$, whereas the Chartists are active with a probability  $p^t_C = 0.5$. This because the interest of  Chartists in purchasing or selling Bitcoins is higher than that of Random traders. Random traders issue orders to satisfy their needs, whereas Chartists issue orders for speculative reasons, study carefully the price variation over time and are readier to place orders. Note that active Chartists actually place orders only if the price variation is above a given threshold.

\section{Number of Traders}\label{sec:App.3}
One of the most attractive property of Bitcoin is to provide quasi-anonymous transactions, so knowing the number of traders in the real market is very difficult.
The Bitcoin addresses used for the transactions are known, but a user can have, and typically has, more than one address. 

At the moment of writing (last quarter of 2015) we had three figures:

\begin{enumerate}
\item the massive and transparent ledger of every Bitcoin transaction was generated starting since January 3, 2009 presumably by the inventor of the Bitcoin system itself, Nakamoto and so in January 2009 there was only one person owning Bitcoins; 
\item 
\begin{quote}
\small{"according to rough estimates, 280,000 people owned Bitcoins at the end of 2013"} (http://www.whoishostingthis.com/blog/2014/03/03/who-owns-all-the-bitcoins/);
\end{quote}
\item 

\begin{quote}
\small{on April 22nd, 2014 the total number of holders was estimated equal to 1.0 million} (https://Bitcointalk.org/index.php?topic=316297.0).
\end{quote}
\end{enumerate}

In addition to these figures, we have other data related to the period between January 2009 and September 2010. These data were extracted from an analysis on the daily number of downloads of the official Bitcoin software client from the SourceForge platform (http://sourceforge.net/projects/bitcoin).

In the period just mentioned, the Bitcoin network began to spread and Bitcoin had no monetary value. So, for this period we assumed the number of downloads of the official Bitcoin software client equal to the number of traders in the market. We made the assumption that a person who downloads the Bitcoin software is mainly interested to use it to mine Bitcoins.
So, we extracted two figures from this data: the total number of downloads on May 1st, 2010 equal to 2,769, and the total number of downloads on September 1st, 2010,  equal to 30,589, and we set the number of downloads equal to the number of traders.

So, we computed the number of traders in the market fitting the curve $N_T$ through the five figures available:
\begin{enumerate}
\item 1 people owned Bitcoins on January 2009. He was Satoshi Nakamoto;
\item 2769  people had downloaded Bitcoin mining software on May 2010;
\item 30589  people had downloaded Bitcoin mining software on September 2010;
\item 280,000 people owned Bitcoins at the end of 2013;
\item 1000000 people owned Bitcoins on April 2014.
\end{enumerate}

The fitting curve of the number of traders $N_T$ is defined by using a general exponential model:

\begin{equation}\label{N_T}
N_T(t) =  a*e^{(b*(608+t))}
\end{equation}

where  $a= 2624$, $b=0.002971$.

\begin{figure}[!ht]
\centering
\subfigure[]{
\includegraphics[width=0.45\textwidth]{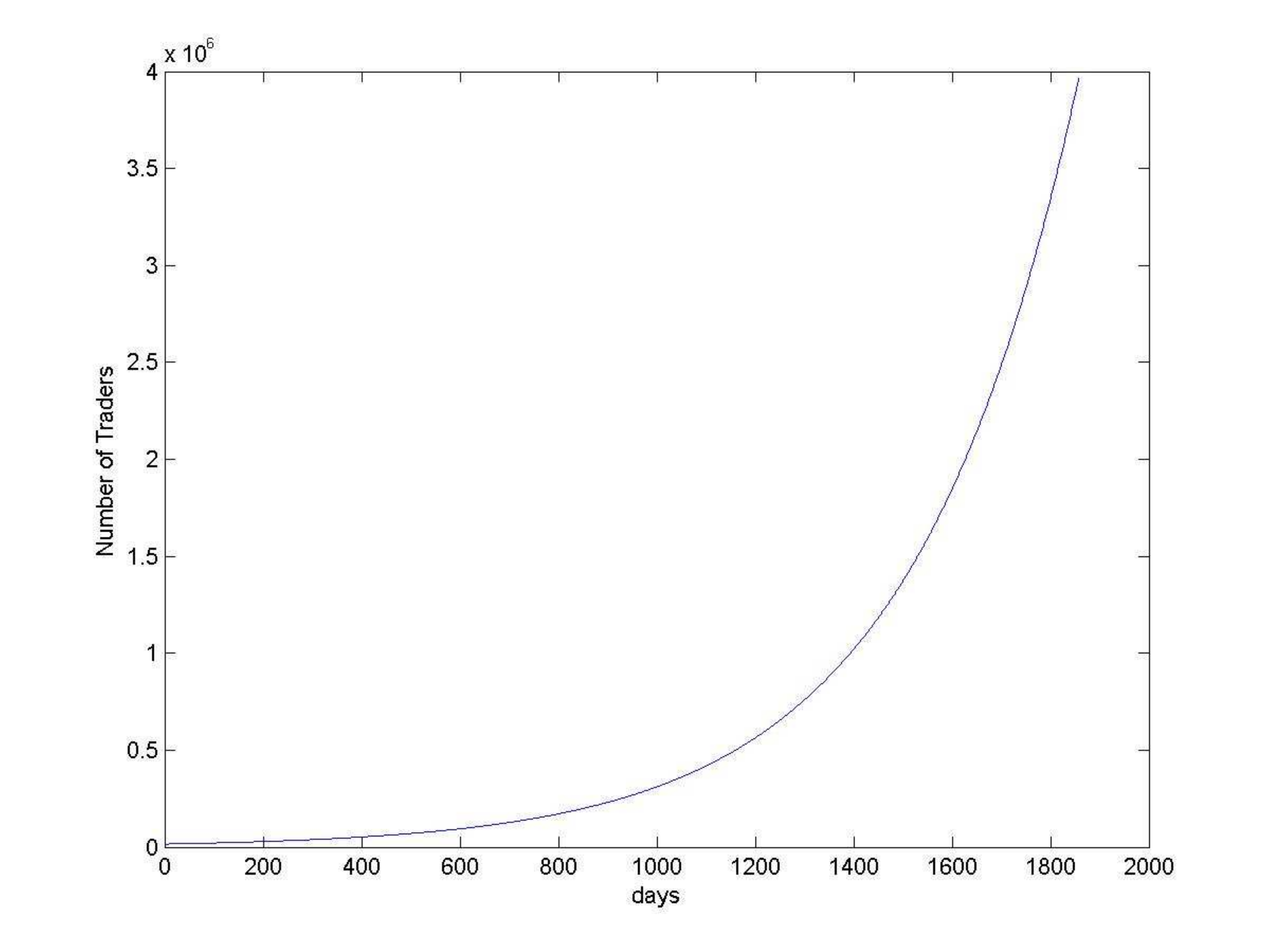}}
\hspace{7mm}
\subfigure[]{
\includegraphics[width=0.45\textwidth]{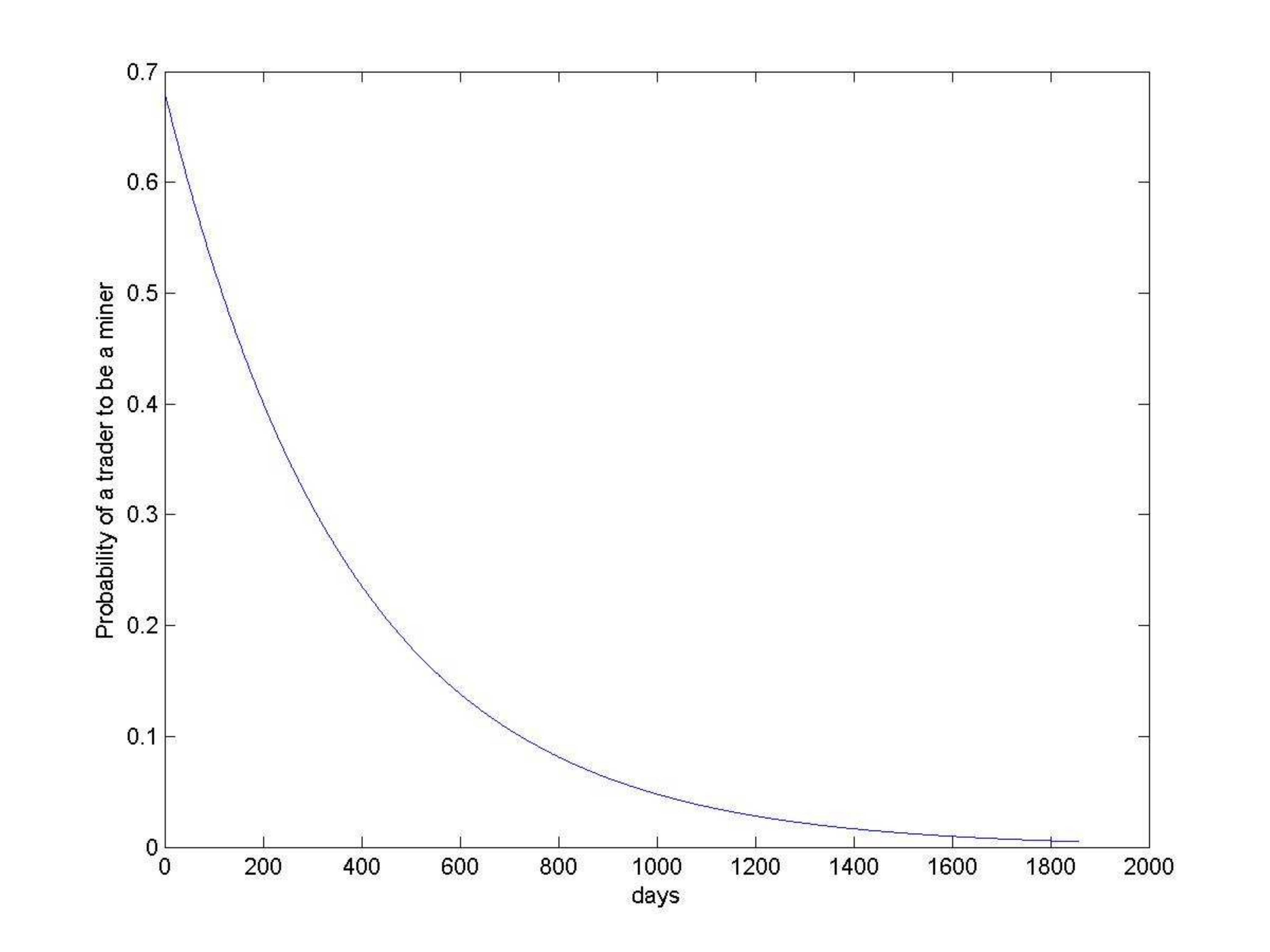}}
\hspace{7mm}
\subfigure[]{
\includegraphics[width=0.45\textwidth]{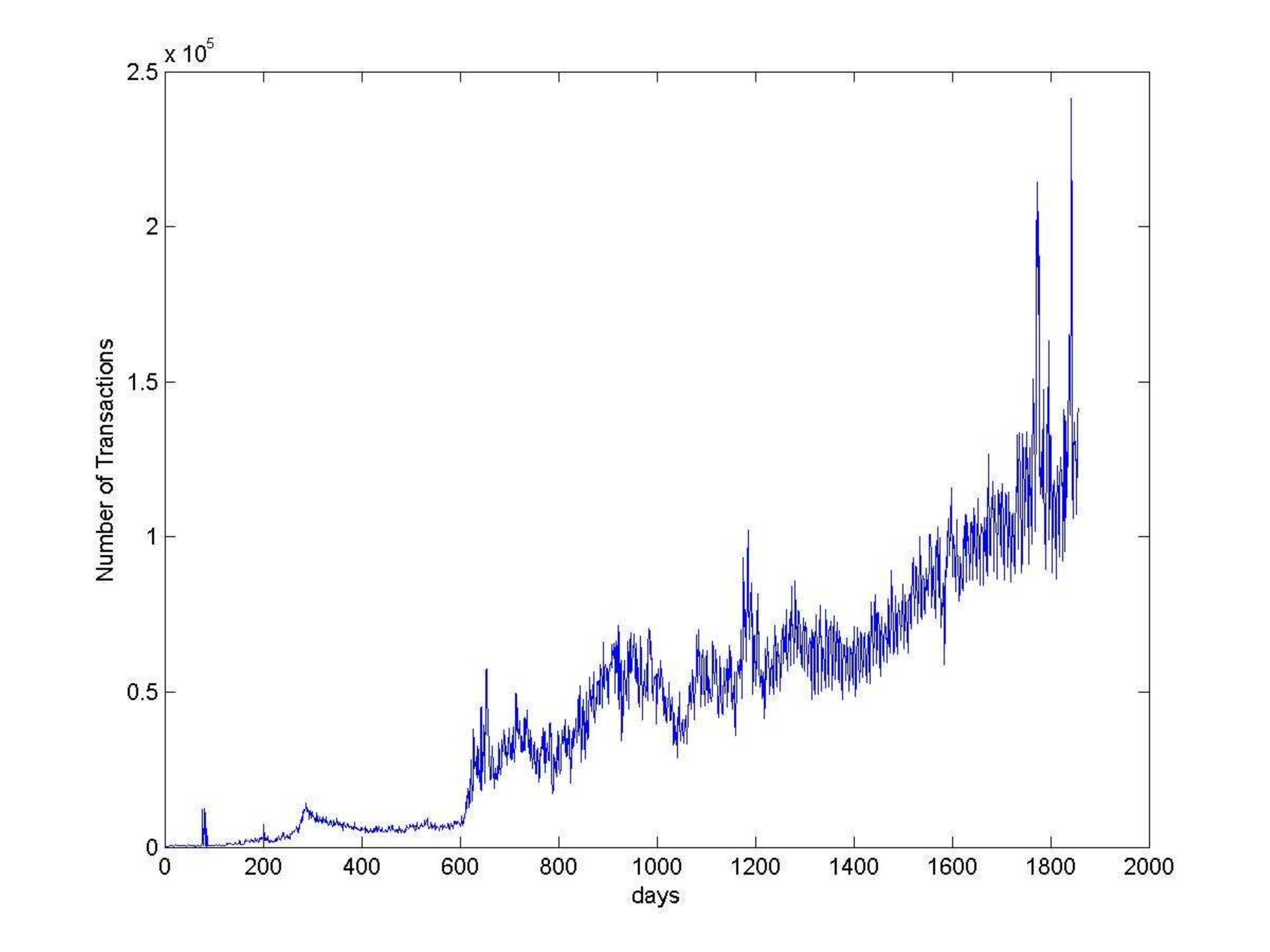}}
\caption{Fitting curve (a) of $N_T(x)$, (b) of the probability of a trader to be a Miner, and (c) number of real Bitcoin transactions. \label{fig:tradProbBeMin}}
\end{figure}

Fig. \ref{fig:tradProbBeMin} (a) show the fitting curve and how the number of traders increases over time.

\section{Probability of a trader to belong to a specific traders' population}\label{sec:App.4}
At first, users in the Bitcoin network were mainly miners. Subsequently, when the Bitcoin network started to grow and Bitcoins started to acquire a monetary value, new users entered the network. Most of these users did not mine, but simply traded Bitcoins. They are represented in the model as Random traders and Chartists.

Since the percentages of the different trader populations is not known, to compute the probability of a trader to be a Miner we performed an analysis of the Bitcoin Blockchain. We analysed the Blockchain until May 1st, 2010 and then until September 1st, 2010.
Each block in the Blockchain contains a list of validated transactions. Each transaction has in input all the addresses containing the amount of Bitcoins to  transfer, and in output all the addresses that receive the Bitcoins. Users can use multiple addresses.

We assumed, as in \cite{Garcia}, that the input addresses that send Bitcoins to the same output address must belong to an unique owner, since to proceed with the transaction it is necessary to know the private keys of all input addresses.

In addition, we considered the input addresses that transfer more than 20 Bitcoins owned by the same owner of the corresponding outputs. In fact, in the period from May 1st, 2010, to September 1st, 2010 Bitcoins had no monetary value, and so it is acceptable to consider that Miners exchanged only small amounts of Bitcoins to test the operation of the network. Of course, this last assumption can be valid only for the period under study.

With these assumptions, we found 46,005 unique addresses on May 1st, 2010 and 82,294 unique addresses on September 1st, 2010. With further analysis, we found that on May 1st, 2010, out of 46,005 addresses, 43,389 were addresses that mined, and on September 1st, 2010, out of 82,294 addresses, 55,974 were addresses that mined, so we computed the probability of an user to be a miner. This probability is equal to 0.94\% on May 1st, 2010 and equal to 0.68\% on September 1st, 2010.

Using these two figures we computed a fitting curve of the probability of an user to be a miner $p_M$. Again, it is defined by using a general exponential model:

\begin{equation}\label{p_M}
p_M(t) =  a*e^{(b*t)}
\end{equation}

where  $a=  0.9425$, $b=-0.002654$.

Fig. \ref{fig:tradProbBeMin} (b) shows the fitting curve and how this probability decreases over time.
We see that the probability of a trader to be a Miner decreases over time, going about from 0.38 to less than 0.01.
Of course, defining this probability using a fitting curve computed from only two points is questionable. In the followings, we made some considerations to validate the adoption of this curve.

At first, the number of Bitcoin transactions was low because in the market there were mainly miners. Over time, as Bitcoin was acquiring monetary value, the number of users interested in exchanging Bitcoins increased. So, while the percentage of Miners in the market was decreasing, also due to the increasing difficult to mine Bitcoins, the percentage of Random traders $p_R$ and Chartists $p_C$ greatly increased, according to the growth of the number of transactions, which slowly rose until a peak on May 2012 and then at the end of the 2013 (see figure \ref{fig:tradProbBeMin} (c)). .

We assumed that the random traders to chartist ratio is 7/3, meaning that 30\% of traders who are not miners are speculators, whereas the remaining 70\% are non-speculative investors. These figures are consistent with recent data obtained for the foreign exchange market \cite{Dick2013}.
The probabilities of non-miners to be a random trader or a chartist, $p_R$ and $p_C$ respectively, are defined as a function of $p_M$, respectively as:

\begin{equation}\label{p_R}
p_R=0.7(1-p_M)
\end{equation}

and 

\begin{equation}\label{p_C}
p_C=0.3(1-p_M)
\end{equation}

With these probabilities, we have at the end of the simulations a number of Miners equal to about 1000, corresponding to 100,000 miners in the real world. This is in agreement with what an Australian bitcoin miner, Andrew Geyl, estimated (see web site http://bravenewcoin.com/news/number-of-bitcoin-miners-far-higher-than-popular-estimates/).

\end{document}